\documentclass[reprint, longbibliography]{revtex4-1}

\usepackage[utf8]{inputenc}

\usepackage[lining,semibold]{libertine} 
\usepackage{amsthm}
\usepackage[libertine, cmintegrals, bigdelims, vvarbb]{newtxmath}

\usepackage{amsmath}
\usepackage{amsfonts}
\usepackage{mathrsfs}
\usepackage{gensymb}
\usepackage{bbm}
\usepackage{dsfont}

\usepackage{kbordermatrix}
\renewcommand{\kbldelim}{(}
\renewcommand{\kbrdelim}{)}

\usepackage{chemformula}
\usepackage[caption=false]{subfig}

\usepackage{soul}
\usepackage{xcolor}

\theoremstyle{definition} 
\newtheorem{example}{Example}

\usepackage{scalerel} 

\makeatletter
\def\maketag@@@#1{\hbox{\m@th\normalfont\normalsize#1}}
\makeatother

\definecolor{webgreen}{rgb}{0,.5,0}
\definecolor{webbrown}{rgb}{.6,0,0}
\definecolor{grigio}{rgb}{.85,.85,.85} 
\definecolor{RoyalBlue}{rgb}{0.0, 0.14, 0.4}
\definecolor{skyblue1}{rgb}{0.45,0.62,0.81}
\definecolor{skyblue2}{rgb}{0.2,0.39,0.64}
\definecolor{skyblue3}{rgb}{0.13,0.29,0.53}
\definecolor{scarlet1}{rgb}{0.93,0.16,0.16}
\definecolor{scarlet2}{rgb}{0.8,0,0}
\definecolor{scarlet3}{rgb}{0.64,0,0}

\definecolor{g}{gray}{0.50}

\usepackage{hyperref}
\hypersetup{%
    colorlinks=true, linktocpage=true, pdfstartpage=1, pdfstartview=FitV,%
    breaklinks=true, pdfpagemode=UseNone, pageanchor=true, pdfpagemode=UseOutlines,%
    plainpages=false, bookmarksnumbered, bookmarksopen=true, bookmarksopenlevel=1,%
    hypertexnames=true, pdfhighlight=/O,
    urlcolor=webbrown, linkcolor=RoyalBlue, citecolor=webgreen, 
    pdftitle={},%
    pdfauthor={Francesco Avanzini},%
    pdfsubject={},%
    pdfkeywords={},%
    pdfcreator={pdfLaTeX},%
    pdfproducer={LaTeX REVTeX}%
}

\DeclareMathAlphabet{\mathpzc}{OT1}{pzc}{m}{it}

\newcommand{\norm}[1]{\lVert#1\rVert}

\newcommand{\steady}[1]{\overline{#1}}

\newcommand{\chemspecies}{\alpha}
\newcommand{\chemspeciesB}{\beta}

\newcommand{\volume}{V}

\newcommand{\setchemspecies}{Z}

\newcommand{\setinternal}{X}
\newcommand{\setchemostatted}{Y}
\newcommand{\setforce}{{Y_f}}
\newcommand{\setpotential}{{Y_p}}

\newcommand{\elrct}{\rho}
\newcommand{\setelrct}{\mathcal R}

\newcommand{\conc}{{\boldsymbol{z}}}

\newcommand{\intconc}{x}
\newcommand{\exconc}{y}

\newcommand{\activitycoeff}{\gamma}
\newcommand{\activitycoeffvec}{{\boldsymbol\gamma}}

\newcommand{\matS}{\mathbb S}
\newcommand{\matSX}{\mathbb S{}^\setinternal}
\newcommand{\matSY}{\mathbb S{}^\setchemostatted}
\newcommand{\matSYf}{\mathbb S{}^\setforce}
\newcommand{\matSYp}{\mathbb S{}^\setpotential}
\newcommand{\effmatSY}{\hat{\mathbb S}{}^\setchemostatted}

\newcommand{\curr}{j}
\newcommand{\excurr}{I}
\newcommand{\excurrY}{{I}^{\setchemostatted}}
\newcommand{\excurrYf}{{I}^{\setforce}}
\newcommand{\excurrYfss}{\steady{\boldsymbol{I}}{}^{\setforce}}

\newcommand{\moieties}{m}

\newcommand{\freerct}{\Delta_\elrct G}
\newcommand{\freeenergy}{G}
\newcommand{\idfreeenergy}{G^{\text{id}}}
\newcommand{\nidfreeenergy}{G^{\text{in}}}

\newcommand{\potenergy}{\nidfreeenergy}
\newcommand{\potential}{g^{\text{in}}}
\newcommand{\fluxexmolarpotenergy}{{\dot{g}}_{\setchemostatted}^{\text{in}}}

\newcommand{\idchempotential}{\mu^{\text{id}}}
\newcommand{\idchempotentialY}{\boldsymbol{\mu}^{\text{id}}_\setchemostatted}
\newcommand{\chempotential}{{\mu}}
\newcommand{\stchempotential}{\mu^{\circ}}

\newcommand{\chargedensity}{Q}

\newcommand{\parameter}{{\boldsymbol{e}}{}}
\newcommand{\scalarparameter}{{e}{}}

\newcommand{\ik}{{k}^{\text{id}}}
\newcommand{\nik}{{k}}
\newcommand{\constant}{\kappa}

\newcommand{\conslaw}{\ell}
\newcommand{\consquantity}{L}
\newcommand{\conslawindex}{\lambda}
\newcommand{\brokenindex}{{\lambda_b}}
\newcommand{\unbrokenindex}{{\lambda_u}}

\newcommand{\matLb}{\mathbb L^{{b}}}
\newcommand{\matLbX}{\mathbb L^{{b}}_{\setinternal}}

\newcommand{\matLbYp}{\mathbb L^{{b}}_{\setpotential}}
\newcommand{\matLbYf}{\mathbb L^{{b}}_{\setforce}}

\newcommand{\cycle}{c}
\newcommand{\cycleindex}{\xi}
\newcommand{\internalindex}{\iota}
\newcommand{\emergentindex}{\epsilon}
\newcommand{\coeffcycle}{\psi}

\newcommand{\nelrct}{|\elrct|}
\newcommand{\nchemostatted}{|\setchemostatted|}
\newcommand{\nforce}{|\setforce|}
\newcommand{\ncycle}{|\internalindex|}
\newcommand{\nemergent}{|\emergentindex|}
\newcommand{\nchemspecies}{|\chemspecies|}
\newcommand{\nconslaw}{|\conslawindex|}
\newcommand{\nbroken}{|\brokenindex|}

\newcommand{\epr}{\dot{\Sigma}}

\newcommand{\statefun}{\Psi}
\newcommand{\molarstatefunscalar}{\mathcal F}
\newcommand{\molarstatefun}{\boldsymbol{\mathcal F}}
\newcommand{\molarstatefunY}{\boldsymbol{\mathcal F}_\setchemostatted}
\newcommand{\molarstatefunYf}{\boldsymbol{\mathcal F}_\setforce}

\newcommand{\semigrand}{\mathcal G}

\newcommand{\ncwr}{\dot{w}_{\text{nc}}}
\newcommand{\ncwrss}{\steady{\dot{w}}_{\text{nc}}}
\newcommand{\dw}{\dot{w}_{\text{driv}}}
\newcommand{\cdw}{\dot{w}_{\text{driv}}^{\text{ch}}}
\newcommand{\edw}{\dot{w}_{\text{driv}}^{\text{in}}}

\newcommand{\relentropy}{\mathcal L}

\newcommand{\manifold}[1]{\Omega\left(#1\right)}

\newcommand{\ncforcescalar}{\mathcal F}
\newcommand{\ncforce}{\boldsymbol{\mathcal F}_{\setforce}}

\newcommand{\matW}{\mathbb W}
\newcommand{\matWY}{\mathbb W^{\setchemostatted}}

\newcommand{\dt}{\mathrm d_t}

\newcommand{\eq}{\text{eq}}
\newcommand{\minl}{\text{min}}

\newcommand{\muteindex}{i}
\newcommand{\setvariable}{i}
\newcommand{\coeff}{\mathpzc f}
\newcommand{\one}{\mathbb 1}

\newcommand{\compone}{{{(1)}}}
\newcommand{\comptwo}{{{(2)}}}
\newcommand{\compthr}{{{(3)}}}

\newcommand{\costone}{\kappa}
\newcommand{\costtwo}{h}

\begin{document}

\title{Nonequilibrium Thermodynamics of Non-Ideal Chemical Reaction Networks}
\newcommand\unilu{\affiliation{Complex Systems and Statistical Mechanics, Department of Physics and Materials Science, University of Luxembourg, L-1511 Luxembourg}}
\author{Francesco Avanzini}
\email{francesco.avanzini@uni.lu}
\unilu
\author{Emanuele Penocchio}
\email{emanuele.penocchio@uni.lu}
\unilu
\author{Gianmaria Falasco}
\email{gianmaria.falasco@uni.lu}
\unilu
\author{Massimiliano Esposito}
\email{massimiliano.esposito@uni.lu}
\unilu

\date{\today}

\begin{abstract}
All current formulations of nonequilibrium thermodynamics of open chemical reaction networks rely on the assumption of non-interacting species.
We develop a general theory which accounts for interactions between chemical species within a mean-field approach using activity coefficients. 
Thermodynamic consistency requires that rate equations do not obey to standard mass-action kinetics, but account for the interactions with concentration dependent kinetic constants.
Many features of the ideal formulations are recovered.
Crucially, the thermodynamic potential and the forces driving non-ideal chemical systems out of equilibrium are identified.
Our theory is general and holds for any mean-field expression of the interactions leading to lower bounded free energies.
\end{abstract}

\maketitle


\section{Introduction}
Thermodynamics studies the interconversions of energy.
It was originally formulated as an equilibrium theory in the 19th century.
Phenomenological extensions close to equilibrium, in the so-called linear regime, were introduced in the first half of the 20th century~\cite{Prigogine1961, Groot1984}.
The earliest formulations beyond the linear regime were developed for stochastic chemical reactions in the second half of the 20th century~\cite{Hill1977, Schnakenberg1976, Jiuli1984, Mou1986, Ross1988}.
But it is only in the 21st century that stochastic thermodynamics was systematized to characterize dissipative processes occurring arbitrarily far from equilibrium and their fluctuations~\cite{Jarzynski2011, Seifert2012, VanDenBroeck2015}.
In recent years, building on the work of Prigogine and co-workers~\cite{Prigogine2015}, stochastic thermodynamics was formulated for stochastic~\cite{Gaspard2004, Schmiedl2007, Rao2018b} and extended to deterministic~\cite{Qian2005, Polettini2014, Rao2016} chemical reaction networks (CRNs) maintained in a nonequilibrium regime with particle reservoirs called chemostats.

Nonequilibrium thermodynamics of CRNs is a powerful tool to analyze chemical complexity.
It has been used for reaction diffusion systems to quantify the energetic cost of creating patterns~\cite{Falasco2018a}, sustaining chemical waves~\cite{ Avanzini2019a} and powering a chemical cloaking device~\cite{Avanzini2020a}.
The growth process of macromolecules like copolymers~\cite{Andrieux2008, Blokhuis2017, Gaspard2020a} and biomolecules~\cite{Rao2015} has been characterized.
Also applications to chaotic CRNs have been considered~\cite{Gaspard2020b}.
Recently, the connections of the theory to information geometry has been investigated~\cite{Yoshimura2020}.
Furthermore, different strategies to maintain nonequilibrium regimes, like the use of finite chemostats~\cite{Fritz2020} and serial transfers in closed reactors~\cite{Blokhuis2018}, have been explored.

All current formulations of nonequilibrium thermodynamics of CRNs are however based on the assumption that the chemical species do not interact, except via chemical reactions.
The purpose of this paper is to generalize nonequilibrium thermodynamics to non-ideal CRNs described by deterministic rate equations.
Interactions are treated with a mean-field approach using the same activity coefficients introduced in equilibrium thermodynamics~\cite{Poniewierski2012}, but expressed in terms of nonequilibrium concentrations.
Exploiting the local detailed balance assumption, we impose thermodynamic consistency on the dynamics.
The standard mass-action kinetics has to be modified introducing concentration dependent kinetic constants to account for the effects of interactions as already recognized in the literature~\cite{Boudart1968, Laidler1987, Butt1999}.

Crucially, we generalize the decomposition of the entropy production developed in Ref.~\onlinecite{Rao2018b} to non-ideal CRNs.
This allows us to determine the thermodynamic potential and the forces driving non-ideal chemical systems out of equilibrium.
The expression of the forces is remarkably similar to the corresponding one for ideal CRNs since all the effects due to the interactions are hidden in the activity coefficients.
The thermodynamic potential still acts as a Lyapunov function in detailed balanced systems, and is always lower bounded by its equilibrium value.
However, the difference between the nonequilibrium and the equilibrium value of the potential cannot be solely expressed in terms of a relative entropy as for ideal CRNs.
Furthermore, the exchange currents controlling the chemostatted species cannot be defined only dynamically as in ideal CRNs, but  they must account for the interactions via the activity coefficients.

To develop our theory in a self-contained way, we proceed as follows.
In Sec.~\ref{sec_crns} we discuss the network theory of chemical reactions.
After introducing the basic notation in Subs~\ref{sec_notation}, we examine the dynamics in Subs.~\ref{sub_dyn}.
We then define conservation laws and cycles in Subs.~\ref{sub_cons_law} and~\ref{sub_cycles}, respectively.
In Subs.~\ref{sub_equilibrium}, we identify under which conditions the existence of equilibrium states is granted, namely, CRNs are detailed balanced. 
We develop our thermodynamic theory in Sec.~\ref{sec_thermo}.
After introducing the general setup in Subs.~\ref{sec_set_up_thermo}, we connect the dynamics to the thermodynamics using the local detailed balance assumption in Subs.~\ref{sub_ldb}. 
The expressions for the exchange currents controlling the chemostatted species are derived in Subs.~\ref{sub_chemostat}.
We then make use of conservation laws to decompose the entropy production rate in Subs.~\ref{sub_epr_decomposition}.
At steady state, the entropy production has a second physically meaningful decomposition based on the cycles as we show in  Subs.~\ref{sub_ss_entropy}.
The properties of the thermodynamic potential are discussed in Subs.~\ref{sub_bound}, where we derive its lower bound, and in Subs.~\ref{sub_relative_entropy}, where we show that it cannot be written as its equilibrium value plus a relative entropy as in ideal systems.

Throughout the manuscript, we use the CRN~\eqref{eq_crn_example} to illustrate our results.
This example represents a catalytic process power by a proton transfer in a system where the relevant interactions are the electrostatic interactions between charged species.
We summarize our results and discuss their implications in Sec.~\ref{sec_conclusion}. 

This work should be particularly relevant for energetic considerations in biosystems, where many metabolic processes involve storing energy in ion gradients across membranes~\cite{voet2010}, and electrochemical systems, where CRNs are coupled to electronic circuits~\cite{Bard2001}.


\section{Non-ideal Chemical Reaction Networks\label{sec_crns}}

\subsection{Setup\label{sec_notation}}
We consider systems composed of interacting chemical species, identified by the label $\chemspecies\in\setchemspecies$, homogeneously distributed in a constant volume $\volume$. 
Our description can treat gas phases and dilute solutions where the volume of the solution is overwhelmingly dominated by the solvent. 
The chemical species undergo elementary chemical reactions~\cite{Svehla1993}, identified by the index $\elrct\in\setelrct$,
\begin{equation}
\boldsymbol \chemspecies\cdot \boldsymbol \nu_{+\elrct} \ch{<=>[ $+\elrct$ ][ $-\elrct$ ]} \boldsymbol \chemspecies\cdot \boldsymbol \nu_{-\elrct}\,,
\label{eq_elementary_reaction}
\end{equation}
with $\boldsymbol \chemspecies =(\dots,\chemspecies,\dots)^\intercal$ the vector of chemical species and $\boldsymbol \nu_{\pm\elrct}$ the vector of stoichiometric coefficients of the forward/backward reaction $\pm\elrct$.
The set of chemical reactions~\eqref{eq_elementary_reaction} defines the CRN.
Our framework considers open CRNs. 
We split the set of all the species $\setchemspecies$ into two disjoint subsets: the internal species~$\setinternal$ and the chemostatted species~$\setchemostatted$. 
The former undergo only the chemical reactions~\eqref{eq_elementary_reaction}. 
The latter too, but they are also externally exchanged.
We will discuss the dynamical implications and the thermodynamic meaning of the chemostatting procedure in Subs.~\ref{sub_dyn} and Subs.~\ref{sub_chemostat}, respectively.

\subsection{Dynamics\label{sub_dyn}}
The state of deterministic CRNs with constant volume is specified by the concentration vector $\conc(t)=(\dots, [\chemspecies](t),\dots)^\intercal$. 
Its dynamics follows  the rate equation
\begin{equation}
\dt \conc(t)=\matS \boldsymbol \curr( \conc(t))+\boldsymbol \excurr(t)\,,
\label{eq_dynamics_crns}
\end{equation}
where we introduced the stoichiometric matrix $\matS$, the current vector $ \boldsymbol \curr( \conc)$ and the exchange current vector $\boldsymbol\excurr(t)$.
The first term on the r.h.s. of Eq.~\eqref{eq_dynamics_crns}, i.e., $\matS \boldsymbol \curr(\conc(t))$, accounts for the concentration changes due to the chemical reactions~\eqref{eq_elementary_reaction}.
The second term, i.e., $\boldsymbol \excurr(t)$, accounts for the external matter flows.

The stoichiometric matrix $\matS$ codifies the topology of the CRN.
Each $\elrct$ column ${\matS}_\elrct$ specifies the net variation of the number of molecules for each species undergoing the $\elrct$ elementary reaction~\eqref{eq_elementary_reaction}, ${\matS}_\elrct =  \boldsymbol\nu_{-\elrct} - \boldsymbol\nu_{+\elrct} $.
The current vector $\boldsymbol \curr(\conc)=(\dots,\curr^\elrct(\conc),\dots)^\intercal$ specifies the net reaction current for every $\elrct$ reaction as the difference between the forward $\curr^{+\elrct}(\conc)$ and backward reaction flux $\curr^{-\elrct}(\conc)$:
\begin{equation}
\curr^{\elrct}(\conc)=\curr^{+\elrct}(\conc) - \curr^{-\elrct}(\conc)\,.
\label{eq_net_current}
\end{equation}
In case of ideal CRNs, the fluxes $\curr^{\pm\elrct}(\conc)$ are expressed in terms of mass-action kinetics~\cite{Groot1984, Pekar2005, Laidler1987}:
\begin{equation}
\curr^{\pm\elrct}(\conc) = \ik_{\pm\elrct}\conc^{\boldsymbol \nu_{\pm\elrct}}\,,
\label{eq_mass_action}
\end{equation}
where $\ik_{\pm\elrct}$ are the kinetic constants of the forward/backward reaction $\pm\elrct$ and we used the following notation $\boldsymbol a^{\boldsymbol b} = \prod_\muteindex a_\muteindex^{b_\muteindex}$.
In case of non-ideal CRNs, mass-action kinetics is not thermodynamically consistent as we show in Subs.~\ref{sub_ldb}.
To take into account the effects of the interactions, we assume that the kinetic constants may depend on the concentrations $\big\{\nik_{\pm\elrct}(\conc)\big\}$ as widely acknowledged in the literature~\cite{Boudart1968, Laidler1987, Butt1999}.
Thus, the reaction fluxes are given by the general expressions
\begin{equation}
\curr^{\pm\elrct}(\conc) = \nik_{\pm\elrct}(\conc)\conc^{\boldsymbol \nu_{\pm\elrct}}\,.
\label{eq_no_mass_action}
\end{equation}
The specific $\conc$-dependence of $\{\nik_{\pm\elrct}(\conc)\}$ derives from the particular model used to describe the interactions.
We develop our theory only assuming that $\nik_{\pm\elrct}(\conc)$ satisfy the conditions introduced in Subs.~\ref{sub_equilibrium} and Subs.~\ref{sub_ldb} to be thermodynamically consistent.

The exchange current vector $\boldsymbol\excurr(t)$ specifies the external matter flows.
It has null entries for the internal species, i.e., $I^{\chemspecies}(t)=0$ for $\chemspecies \in\setinternal$: the concentration of the internal species changes only because of the chemical reactions~\eqref{eq_elementary_reaction} by definition.
The entries of $\boldsymbol\excurr(t)$ for the chemostatted species,  i.e., $I^{\chemspecies}(t)$ for $\chemspecies \in\setchemostatted$, are derived in Subs.~\ref{sub_chemostat} because the thermodynamic meaning of the chemostatting procedure must be taken into account.

We can apply the splitting of the chemical species into internal and chemostatted ones to the stoichiometric matrix,
\begin{equation}
\matS=\begin{pmatrix}
\matSX \\ 
\matSY\\
\end{pmatrix}\,,
\end{equation}
and the concentration vector $\conc= (\boldsymbol \intconc,\boldsymbol\exconc)$. Analogously, the rate equation~\eqref{eq_dynamics_crns} becomes
\begin{align}
&\dt \boldsymbol \intconc(t)=\matSX \boldsymbol \curr(\boldsymbol \intconc(t), \boldsymbol \exconc(t))\label{eq_dynamics_crns_X}\,,\\
&\dt \boldsymbol \exconc(t)=\matSY \boldsymbol \curr(\boldsymbol \intconc(t), \boldsymbol \exconc(t)) + \boldsymbol \excurrY(t)\label{eq_dynamics_crns_Y}\,,
\end{align}
with $ \boldsymbol \excurrY(t) = (\dots,\excurr^\chemspecies(t),\dots )^\intercal_{\chemspecies\in\setchemostatted}$ collecting the not null entries of $\boldsymbol \excurr(t)$. Note that the Eqs.~\eqref{eq_dynamics_crns_X} and~\eqref{eq_dynamics_crns_Y} are only a reformulation of Eq.~\eqref{eq_dynamics_crns}.

\begin{example}
We consider the following CRN (see also Fig.~\ref{fig_scheme_ATP}):
\begin{figure}[t]
\centering
\includegraphics[width=1.\columnwidth]{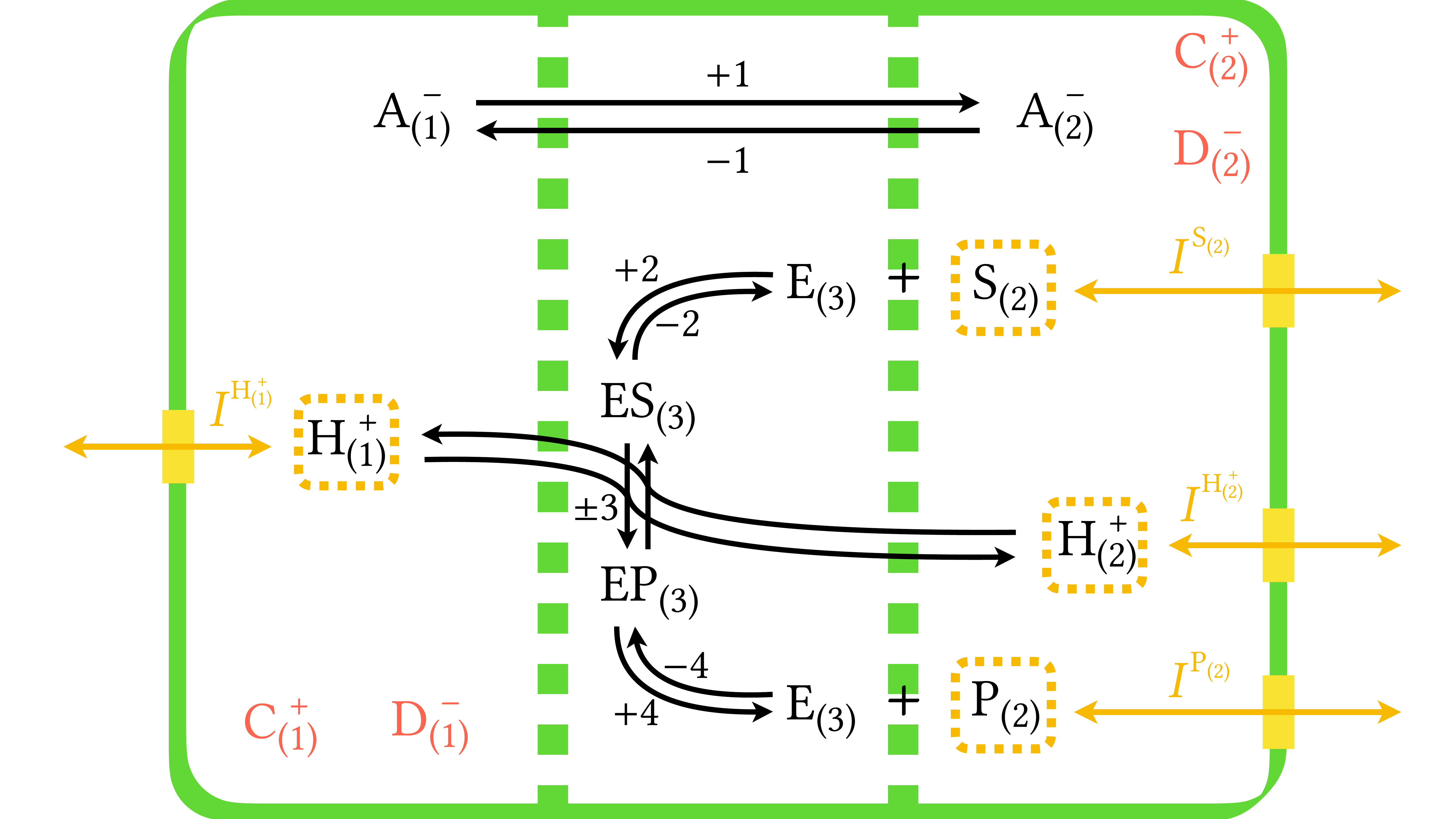}
\caption{Drawing of the CRN~\eqref{eq_crn_example}.
A membrane (green dashed line) divides the system into compartments.
A membrane enzyme $\ch{E}$ binds the substrate $\ch{S}$. 
The interconversion of the substrate $\ch{S}$ into the product $\ch{P}$ is coupled to the proton $\ch{H^{+}}$ transfer from compartment  $\compone$ to compartment $\comptwo$.
In parallel, the anions~\ch{A^{-}} move from compartment  $\compone$ to compartment $\comptwo$.
The $\ch{S}$, $\ch{P}$, $\ch{H^{+}}$ are chemostatted (yellow arrows).
Other cations~\ch{C^{+}} and anions~\ch{D^{-}} are present in compartment $\compone$ and $\comptwo$.}
\label{fig_scheme_ATP}
\end{figure}
\begin{equation}
\begin{split}
\ch{A^{-}_{\compone} &<=>[ ${+1}$ ][ ${-1}$ ] A^{-}_{\comptwo} }\\
\ch{S_{\comptwo} + E_{\compthr}  &<=>[ ${+2}$ ][ ${-2}$ ] ES_{\compthr} }\\
\ch{H^{+}_{\compone} + ES_{\compthr}  &<=>[ ${+3}$ ][ ${-3}$ ] EP_{\compthr} + H^{+}_{\comptwo} }\\
\ch{EP_{\compthr} &<=>[ ${+4}$ ][ ${-4}$ ] E_{\compthr}  + P_{\comptwo}}
\end{split}
\label{eq_crn_example}
\end{equation}
where the superscripts indicate the electric charge and the subscripts indicate the compartment where a chemical species is located.
Here, $\chemspecies \in \setchemspecies = \{\ch{E_{\compthr} },\ch{ES_{\compthr} },\ch{EP_{\compthr} },\ch{A^{-}_{\compone}},\ch{A^{-}_{\comptwo}}, \ch{H^{+}_{\compone}},\ch{H^{+}_{\comptwo}}, \ch{S_{\comptwo}},\ch{P_{\comptwo}}\}$.
The CRN~\eqref{eq_crn_example} represents the transformation of the substrate~\ch{S} into the product~\ch{P} inside compartment~$\comptwo$ catalysed by the membrane enzyme~\ch{E}.
The interconversion of the complex~\ch{ES} into~\ch{EP} on the membrane (compartment~$\compthr$) is coupled to the transfer of protons~\ch{H^{+}} from compartment~$\compone$ to compartment~$\comptwo$.
Independently, anions~\ch{A^{-}} are free to move from compartment~$\compone$ to compartment~$\comptwo$. 
Other cations~\ch{C^{+}} and anions~\ch{D^{-}} are present in compartment~$\compone$ and~$\comptwo$, but they are not involved in any chemical reaction.
For this reason, the species~\ch{C^{+}} and~\ch{D^{-}}  are not included in~$\setchemspecies$.
Because of the coupling between the catalytic reaction and the proton transfer, the CRN~\eqref{eq_crn_example} resembles the working mechanism of the membrane enzyme ATP synthase~\cite{Huxley2000}.
All the protons, the substrate and the product are chemostatted, i.e., $\setchemostatted=\{\ch{H^{+}_{\compone}},\ch{H^{+}_{\comptwo}}, \ch{S_{\comptwo}},\ch{P_{\comptwo}}\}$ and $\setinternal=\{\ch{E_{\compthr} },\ch{ES_{\compthr} },\ch{EP_{\compthr} },\ch{A^{-}_{\compone}},\ch{A^{-}_{\comptwo}}\}$.
For this kind of systems the strongest interactions are the electrostatic interactions between charged species inside the same compartment.

The stoichiometric matrix of the CRN~\eqref{eq_crn_example} is specified as
\begin{equation}
{\matS}=
 \kbordermatrix{
    & \color{g}1 &\color{g}2&\color{g}3&\color{g}4\cr
    \color{g}\ch{E_{\compthr}}    &0 &-1& 0 & 1\cr
    \color{g}\ch{ES_{\compthr}}   &0 &1&-1&0\cr
    \color{g}\ch{EP_{\compthr}}   &0 &0&1&-1\cr
    \color{g}\ch{A^{-}_{\compone}}   &-1 &0&0&0\cr
    \color{g}\ch{A^{-}_{\comptwo}} &1 &0&0&0\cr
    \color{g}\ch{H^{+}_{\compone}}   &0 &0&-1&0\cr
    \color{g}\ch{H^{+}_{\comptwo}} &0 &0&1&0\cr
    \color{g}\ch{S_{\comptwo}}   &0 &-1&0&0\cr
    \color{g}\ch{P_{\comptwo}}   &0 &0&0&1\cr
  }\,.
  \label{eq_crn_example_matS}
\end{equation}
\qed
\end{example}

\subsection{Conservation Laws\label{sub_cons_law}} 
The linearly independent vectors $\{\boldsymbol \conslaw^\conslawindex\}$ in the cokernel of the stoichiometric matrix 
\begin{equation}
\boldsymbol \conslaw^\conslawindex \cdot \matS=0
\label{eq_conservation_laws}
\end{equation}
are the so-called conservation laws~\cite{Polettini2014,Rao2016}.
All the scalars $\consquantity^\conslawindex(\conc) \equiv \boldsymbol \conslaw^\conslawindex\cdot \conc$ would be conserved quantities if the CRN was closed, namely, $\boldsymbol\excurr(t)=0$. Indeed, $\dt\consquantity^\conslawindex (\conc(t))=  \boldsymbol {\conslaw}^\conslawindex \cdot \matS \boldsymbol \curr(\conc(t)) =0$. 
In closed CRNs the total mass is conserved and, consequently, the set $\{\boldsymbol \conslaw^\conslawindex\}$ is never empty. 

When CRNs are open, some conservation laws do not correspond anymore to conserved quantities. 
Hence, we split the set of conservation laws $\{\boldsymbol \conslaw^\conslawindex\}$ into two disjoint subsets: 
the unbroken conservation laws $\{\boldsymbol \conslaw^\unbrokenindex\}$ and the broken conservation laws $\{\boldsymbol \conslaw^\brokenindex\}$. 
The unbroken conservation laws are the largest subset of conservation laws that can be written with null entries for the chemostatted species, i.e., $ \conslaw^\unbrokenindex_{\chemspecies}=0$ for $\chemspecies\in\setchemostatted$. 
Their corresponding scalar quantities $\consquantity^\unbrokenindex(\conc) = \boldsymbol \conslaw^\unbrokenindex\cdot \conc$ are still conserved:
\begin{equation}
\dt\consquantity^\unbrokenindex (\conc(t))=  \underbrace{\boldsymbol {\conslaw}^\unbrokenindex \cdot \matS \boldsymbol \curr(\conc(t))}_{=0} + \sum_{\chemspecies\in\setchemostatted} \underbrace{\conslaw^\unbrokenindex_{\chemspecies}}_{=0}\excurr^\chemspecies(t) =0\,.
\end{equation}
The broken conservation laws are the other conservation laws $\{\boldsymbol \conslaw^\brokenindex\}= \{\boldsymbol \conslaw^\conslawindex\}\setminus \{\boldsymbol \conslaw^\unbrokenindex\}$. 
Their corresponding scalar quantities $\consquantity^\brokenindex(\conc) = \boldsymbol \conslaw^\brokenindex\cdot \conc$ are, in general, not conserved and they give rise to balance equations:
\begin{equation}
\dt\consquantity^\brokenindex (\conc(t))=  \underbrace{\boldsymbol {\conslaw}^\brokenindex \cdot \matS \boldsymbol \curr(\conc(t))}_{=0} + \sum_{\chemspecies\in\setchemostatted} \underbrace{\conslaw^\brokenindex_{\chemspecies}}_{\neq0}\excurr^\chemspecies(t) \neq0\,,
\end{equation}
where the variations of $\consquantity^\brokenindex (\conc(t))$ are due to the matter exchanged through the currents $\{\excurr^\chemspecies(t)\}_{\chemspecies\in\setchemostatted}$.
In open CRNs the total mass is not conserved and, consequently, the set $\{\boldsymbol \conslaw^\brokenindex\}$ is never empty.

Chemostatting a species does not always break a conservation law~\cite{Rao2016, Rao2018b, Falasco2018a, Rao2018a}.
We thus distinguish the species $\setpotential\subseteq \setchemostatted$ that break the conservation laws from the others $\setforce= \setchemostatted \setminus \setpotential$. 
This classification is not unique and different choices have different physical meanings. 
It should be motivated by the physical role played by each chemostatted species as it will become clear in Subs.~\ref{sub_epr_decomposition}. 
We apply the same splitting to the stoichiometric matrix $\matSY$ and the matrix $\matLb$ whose rows are the broken conservation laws $\{\boldsymbol {\conslaw}^\brokenindex\}$:
\begin{align}
&\matSY = \begin{pmatrix}
\matSYf\\
\matSYp
\end{pmatrix}\,,
&\matLb=\begin{pmatrix}
\matLbX, \matLbYf, \matLbYp
\end{pmatrix}\,,
\end{align}
where $\matLbX$, $\matLbYf$, and $\matLbYp$ have $\{{\conslaw}^\brokenindex_\chemspecies\}_{\chemspecies\in\setinternal}$, $\{{\conslaw}^\brokenindex_\chemspecies\}_{\chemspecies\in\setforce}$, and $\{{\conslaw}^\brokenindex_\chemspecies\}_{\chemspecies\in\setpotential}$ as entries, respectively.
Note that the number of $\setpotential$ species is equal to the number of broken conservation laws, by definition. 
The matrix $\matLbYp$ is then square and nonsingular, and so it can be inverted. 
We will exploit this properties in Subs.~\ref{sub_epr_decomposition} where we decompose the entropy production rate.

\begin{example}
Given the stoichiometric matrix~\eqref{eq_crn_example_matS} of the CRN~\eqref{eq_crn_example}, there are five conservation laws,
\begin{equation}
\begin{split}
\boldsymbol \conslaw^{\ch{E}}=
 \kbordermatrix{
     & \cr
    \color{g}\ch{E_{\compthr}}    &1\cr
    \color{g}\ch{ES_{\compthr}}   &1\cr
    \color{g}\ch{EP_{\compthr}}   &1\cr
    \color{g}\ch{A^{-}_{\compone}}   &0\cr
    \color{g}\ch{A^{-}_{\comptwo}} &0\cr
    \color{g}\ch{H^{+}_{\compone}}   &0\cr
    \color{g}\ch{H^{+}_{\comptwo}} &0\cr
    \color{g}\ch{S_{\comptwo}}   &0\cr
    \color{g}\ch{P_{\comptwo}}   &0\cr
  }\,,\text{ }\text{ }\text{ }\text{ }
\boldsymbol \conslaw^{\ch{A}}=&
 \kbordermatrix{
     & \cr
    \color{g}\ch{E_{\compthr}}    &0\cr
    \color{g}\ch{ES_{\compthr}}   &0\cr
    \color{g}\ch{EP_{\compthr}}   &0\cr
    \color{g}\ch{A^{-}_{\compone}}   &1\cr
    \color{g}\ch{A^{-}_{\comptwo}} &1\cr
    \color{g}\ch{H^{+}_{\compone}}   &0\cr
    \color{g}\ch{H^{+}_{\comptwo}} &0\cr
    \color{g}\ch{S_{\comptwo}}   &0\cr
    \color{g}\ch{P_{\comptwo}}   &0\cr
  }\,,\text{ }\text{ }\text{ }\text{ }
 \boldsymbol \conslaw^{\ch{H}}=
 \kbordermatrix{
     & \cr
    \color{g}\ch{E_{\compthr}}    &0\cr
    \color{g}\ch{ES_{\compthr}}   &0\cr
    \color{g}\ch{EP_{\compthr}}   &0\cr
    \color{g}\ch{A^{-}_{\compone}}   &0\cr
    \color{g}\ch{A^{-}_{\comptwo}} &0\cr
    \color{g}\ch{H^{+}_{\compone}}   &1\cr
    \color{g}\ch{H^{+}_{\comptwo}} &1\cr
    \color{g}\ch{S_{\comptwo}}   &0\cr
    \color{g}\ch{P_{\comptwo}}   &0\cr
  }\,,\\
  \boldsymbol \conslaw^{\ch{HS}}=
 \kbordermatrix{
     & \cr
    \color{g}\ch{E_{\compthr}}    &0\cr
    \color{g}\ch{ES_{\compthr}}   &1\cr
    \color{g}\ch{EP_{\compthr}}   &0\cr
    \color{g}\ch{A^{-}_{\compone}}   &0\cr
    \color{g}\ch{A^{-}_{\comptwo}} &0\cr
    \color{g}\ch{H^{+}_{\compone}}   &0\cr
    \color{g}\ch{H^{+}_{\comptwo}} &1\cr
    \color{g}\ch{S_{\comptwo}}   &1\cr
    \color{g}\ch{P_{\comptwo}}   &0\cr
  }\,,&\text{ }\text{ }\text{ }\text{ }
  \boldsymbol \conslaw^{\ch{S}}=
 \kbordermatrix{
     & \cr
    \color{g}\ch{E_{\compthr}}    &0\cr
    \color{g}\ch{ES_{\compthr}}   &1\cr
    \color{g}\ch{EP_{\compthr}}   &1\cr
    \color{g}\ch{A^{-}_{\compone}}   &0\cr
    \color{g}\ch{A^{-}_{\comptwo}} &0\cr
    \color{g}\ch{H^{+}_{\compone}}   &0\cr
    \color{g}\ch{H^{+}_{\comptwo}} &0\cr
    \color{g}\ch{S_{\comptwo}}   &1\cr
    \color{g}\ch{P_{\comptwo}}   &1\cr
  }\,.
\end{split}
\end{equation}
Four of them have a clear physical meaning:
the total concentration of the enzyme is given by $\consquantity^{\ch{E}} = \boldsymbol \conslaw^{\ch{E}}\cdot\conc =  [\ch{E_\compthr}] + [\ch{ES_\compthr}]+[\ch{EP_\compthr}]$,
of the anions by $\consquantity^{\ch{A}} = \boldsymbol \conslaw^{\ch{A}}\cdot\conc = [\ch{A^{-}_\compone}]+[\ch{A^{-}_\comptwo}]$,
of the protons by $\consquantity^{\ch{H}} = \boldsymbol \conslaw^{\ch{H}}\cdot\conc =  [\ch{H^{+}_\compone}]+[\ch{H^{+}_\comptwo}]$,
and of the substrate by $\consquantity^{\ch{S}} = \boldsymbol \conslaw^{\ch{S}}\cdot\conc = [\ch{ES_\compthr}]+[\ch{EP_\compthr}]+[\ch{S_\comptwo}]+[\ch{P_\comptwo}]$.
The conserved quantity $\consquantity^{\ch{HS}}= \boldsymbol \conslaw^{\ch{HS}}\cdot\conc= [\ch{ES_\compthr}]+ [\ch{S_\comptwo}]+[\ch{H^{+}_\comptwo}]$ comes from the coupling between the consumption of the substrate and the proton transfer. 

When the species $\ch{H^{+}_{\compone}}$, $\ch{H^{+}_{\comptwo}}$, $\ch{S_{\comptwo}}$, and $\ch{P_{\comptwo}}$ are chemostatted, the three conservation laws $\boldsymbol \conslaw^{\ch{H}}$, $\boldsymbol \conslaw^{\ch{HS}}$, and $\boldsymbol \conslaw^{\ch{S}}$ are broken.
We identify the set $\setpotential$ as $\setpotential=\{\ch{H^{+}_{\comptwo}},\ch{S_{\comptwo}},\ch{P_{\comptwo}}\}$, but different choices are possible (see the discussion in Appendix A of Ref.~\onlinecite{Avanzini2019a}).
Thus, $\setforce=\setchemostatted\setminus\setpotential=\{\ch{H^{+}_{\compone}}\}$.
The matrix $\matLb$ whose rows are the broken conservation laws is given by
\begin{equation}
\matLb = 
 \kbordermatrix{
    & \color{g}\ch{E_{\compthr}} &\color{g}\ch{ES_{\compthr} }&\color{g}\ch{EP_{\compthr} }&\color{g}\ch{A^{-}_{\compone}}& \color{g}\ch{A^{-}_{\comptwo}}&&\color{g}\ch{H^{+}_{\compone}}&&\color{g}\ch{H^{+}_{\comptwo}}&\color{g}\ch{S_{\comptwo}}&\color{g}\ch{P_{\comptwo}}\cr
    \color{g}\boldsymbol\conslaw^{\ch{H}} &0 &0& 0 & 0&0 &\color{g}\vrule&1& \color{g}\vrule&1 & 0&0\cr
    \color{g}\boldsymbol\conslaw^{\ch{HS}} &0 &1& 0 & 0&0&\color{g}\vrule &0& \color{g}\vrule&1 & 1&0\cr
    \color{g}\boldsymbol\conslaw^{\ch{S}}   &0 &1& 1 & 0&0 &\color{g}\vrule&0& \color{g}\vrule&0 & 1&1\cr
  }\,,
\label{eq_crn_example_matL}
\end{equation}
where the grey vertical lines mark the split of $\matLb$ into $\matLbX$, $\matLbYf$, and $\matLbYp$.
\qed
\end{example}

\subsection{Cycles\label{sub_cycles}}
The linearly independent vectors $\{\boldsymbol \cycle_\internalindex\}$ in the kernel of the stoichiometric matrix
\begin{equation}
\matS \boldsymbol \cycle_\internalindex=0
\label{eq_internal_cycles}
\end{equation}
are the so-called internal cycles~\cite{Polettini2014,Rao2016}.
They represent sequences of reactions that upon completion leave the all concentrations unchanged.
Any linear combination of internal cycles gives a steady-state current vector of closed CRNs, $\dt\conc(t)=\matS\steady{\boldsymbol\curr}=0$ with $\steady{\boldsymbol\curr}=\sum_\internalindex\boldsymbol\cycle_\internalindex\coeffcycle^\internalindex$.

When CRNs are open, the steady-state current vector must 
i) leave the concentrations of the internal species unchanged, i.e., $\matSX\steady{\boldsymbol\curr}=0$, and 
ii) be balanced by the exchanged current vector, i.e., $\matSY\steady{\boldsymbol\curr} + \boldsymbol \excurrY=0$. 
Thus, $\steady{\boldsymbol\curr}\in\ker(\matSX)$ and, consequently, can be written as a linear combination of the right-null eigenvectors
\begin{equation}
\matSX \boldsymbol \cycle_\cycleindex=0\,.
\label{eq_emergent_cycles}
\end{equation}
The set $\{ \boldsymbol \cycle_\cycleindex\}$ includes the internal cycles $\{\boldsymbol \cycle_\internalindex\}$ and, in general, other vectors called emergent cycles $\{\boldsymbol \cycle_\emergentindex\}$.
The steady-state current vector can therefore be written as
\begin{equation}
\steady{\boldsymbol\curr}=\sum_\internalindex\boldsymbol\cycle_\internalindex\coeffcycle^\internalindex +\sum_\emergentindex\boldsymbol\cycle_\emergentindex\coeffcycle^\emergentindex \,.
\label{eq_ss_current}
\end{equation}
Because of the rank-nullity theorem for the stoichiometric matrix $\matS$ and $\matSX$,
every time a species is chemostatted either a conservation law is broken or an emergent cycle arises, namely,
\begin{equation}
\nchemostatted = \nbroken + \nemergent\,,
\end{equation}
with $\nchemostatted$, $\nbroken$ and $\nemergent$ the number of chemostatted species, broken conservation laws and emergent cycles, respectively.
This implies that the number of $\setforce$ species is equal to the number of emergent cycles, i.e., $\nforce = \nemergent$.

\begin{example}
Given the stoichiometric matrix~\eqref{eq_crn_example_matS} of the CRN~\eqref{eq_crn_example}, there is only one emergent cycle and no internal cycles,
\begin{equation}
\boldsymbol \cycle=
 \kbordermatrix{
     & \cr
    \color{g}1    &0\cr
    \color{g}2   &1\cr
    \color{g}3  &1\cr
    \color{g}4   &1\cr
  }\,.
\label{eq_crn_example_emergent_cycle}
\end{equation}
\qed
\end{example}

\subsection{Equilibrium\label{sub_equilibrium}}
The equilibrium steady-state $\conc_\eq$ of the rate equation~\eqref{eq_dynamics_crns}, if it exists, is characterized by vanishing reaction currents,
\begin{equation}
\boldsymbol \curr(\conc_\eq)=0\,.
\label{eq_local_detailed_balance_kinetics}
\end{equation}
This, together with Eq.~\eqref{eq_no_mass_action}, implies
\begin{equation}
\frac{ \nik_{+\elrct}(\conc_\eq)}{ \nik_{-\elrct}(\conc_\eq)}=\conc_\eq^{\matS_\elrct}\,.
\label{eq_local_detailed_balance_kinetics_2}
\end{equation}
If a solution $\conc_\eq$ of Eq.~\eqref{eq_local_detailed_balance_kinetics} (or equivalently of Eq.~\eqref{eq_local_detailed_balance_kinetics_2}) exists, the CRN is said to be \textit{detailed balanced}. 
Closed CRNs must be detailed balanced for thermodynamic consistency. 
It is not granted that open CRNs are detailed balanced. 
It depends on the chemostatting procedure. 

Here, we show that open CRNs are necessarily detailed balanced if the number of chemostatted species is equal to the number of broken conservation laws, i.e., $\setchemostatted = \setpotential$, or, equivalently, there are no emergent cycles.
In Subs.~\ref{sub_epr_decomposition}, we show that equilibrium states can exist also when $\setchemostatted \neq \setpotential$ and there are emergent cycles.
For thermodynamic consistency, namely the local detailed balance condition that will be introduced in Eq.~\eqref{eq_local_detailed_balance_thermo_kinetic_constant},
the product of the forward rate constants along every internal cycle~\eqref{eq_internal_cycles} must equal that of the backward rate constants:
\begin{equation}
\prod_\elrct \left(\frac{\nik_{+\elrct}(\conc)}{\nik_{-\elrct}(\conc)}\right)^{\cycle^\elrct_\internalindex}=1\,,
\end{equation}
where $\cycle^\elrct_\internalindex$ is the $\elrct$ entry of the $\internalindex$ internal cycle. 
This is know as Wegscheider's condition in the framework of ideal CRNs~\cite{Schuster1989}.
Hence, only $\nelrct-\ncycle$ ratios $\nik_{+\elrct}(\conc)/\nik_{-\elrct}(\conc)$ are independent (with $\nelrct$ the number of reactions and $\ncycle$ the number of internal cycles). 
This means that Eq.~\eqref{eq_local_detailed_balance_kinetics_2} imposes  $\nelrct-\ncycle$ constraints. 
This is independent on whether CRNs are closed or open.

In closed CRNs, only $\nchemspecies-\nconslaw$ of the equilibrium concentrations are independent (with $\nchemspecies$ the number of all the species in $\setchemspecies$ and $\nconslaw$ the number of conservation laws). 
The others are fixed by the initial conditions through the conserved quantities:
\begin{equation}
\consquantity^\conslawindex = \boldsymbol \conslaw^{\conslawindex}\cdot \conc(0)= \boldsymbol \conslaw^{\conslawindex}\cdot \conc_{\eq}\,.
\end{equation}
Therefore, Eq.~\eqref{eq_local_detailed_balance_kinetics_2} imposes $\nelrct-\ncycle$ constraints for $\nchemspecies-\nconslaw$ independent variables. Because of the rank-nullity theorem for the stoichiometric matrix $\matS$, 
\begin{equation}
\nelrct-\ncycle=\nchemspecies-\nconslaw\,,
\label{eq_rank_nullity_theorem}
\end{equation}
the existence of a solution to Eq.~\eqref{eq_local_detailed_balance_kinetics_2} is granted.
Closed CRNs must be detailed balanced and hence each solution of Eq.~\eqref{eq_local_detailed_balance_kinetics_2} must be physically meaningful, i.e., $[\chemspecies]_\eq\in\mathbb R_{\geq0}$ $\forall\chemspecies\in\setchemspecies$. 
This imposes constraints on the possible expressions of $\big\{\nik_{\pm\elrct}(\conc)\big\}$.

In open CRNs, $\nconslaw - \nbroken$ concentrations are fixed by the initial conditions through the unbroken conserved quantities:
\begin{equation}
\consquantity^\unbrokenindex = \boldsymbol \conslaw^{\unbrokenindex}\cdot \conc(0)= \boldsymbol \conslaw^{\unbrokenindex}\cdot \conc_{\eq}\,.
\end{equation}
Other $\nchemostatted$ concentrations are constrained by the chemostatting procedure.
The final number of independent concentrations is given by
\begin{equation}
\nchemspecies- (\nconslaw - \nbroken) - \nchemostatted = \nchemspecies-\nconslaw-\nforce= \nelrct-\ncycle - \nforce\, ,
\end{equation}
where we used $\nchemostatted - \nbroken = \nforce$ and Eq.~\eqref{eq_rank_nullity_theorem}.
Therefore, Eq.~\eqref{eq_local_detailed_balance_kinetics_2} imposes $\nelrct-\ncycle$ constraints for $\nelrct-\ncycle - \nforce$ independent variables. 
The existence of a solution is always granted if and only if $\nforce=0$, namely, all the chemostatted species break a conservation law (i.e., $\setchemostatted = \setpotential$).
Equivalently, the existence of a solution is always granted if and only if there are no emergent cycles, $\nemergent= 0$.
The solution is physical meaningful because it must also be a solution of the corresponding closed CRN.

These are sufficient conditions for the existence of an equilibrium state $\conc_\eq$. 
For ideal detailed balanced CRNs, i.e., evolving according to mass-action kinetics~\eqref{eq_mass_action}, there is a unique equilibrium state for every stoichiometric compatibility class $\manifold{\big\{\consquantity^\unbrokenindex\big\}}$, namely, for every manifold in the concentration space characterized by specific and unique values of all the conserved quantities~\cite{Horn1972, Schuster1989}.
For non-ideal CRNs, this property is not granted anymore and there might be more than one equilibrium state for every stoichiometric compatibility class. 
The existence of multiple equilibria in the same stoichiometric compatibility class will become relevant in Subs.~\ref{sub_bound}.

\begin{example}
It is not granted that the open CRN~\eqref{eq_crn_example} is detailed balanced because the set $\setforce$ is not empty, i.e., $\setforce =\{\ch{H^{+}_{\compone}}\}$, and there is the emergent cycle~\eqref{eq_crn_example_emergent_cycle}.
\qed
\end{example}

\section{Thermodynamics\label{sec_thermo}}

\subsection{Setup\label{sec_set_up_thermo}}
Nonequilibrium thermodynamics of CRNs presumes that all degrees of freedom other than concentrations are equilibrated. 
The temperature $T$ is set by a thermal reservoir (e.g., the solvent in dilute solutions), diffusion processes are fast enough to keep  the chemical species homogeneously distributed, and the fields responsible for the interactions, e.g., electrostatic fields, relax instantaneously to their mean-field values.
For dilute solutions, the pressure $p$ is set by the environment of the solution. 
In absence of reactions, the CRN would be an equilibrated mixture.
In this way, thermodynamic state functions can be specified by their equilibrium form but expressed in terms of nonequilibrium concentrations.

We assume that the free energy of non-ideal CRNs, for a given temperature $T$ and volume $\volume$, is a mean-field function $\freeenergy(\conc, \parameter)$ of the concentrations $\conc$ and of some externally controlled parameters $\parameter = (\dots,\scalarparameter_\muteindex,\dots)^\intercal$ that tune the interactions. 
We chose the symbol $\freeenergy$ for the free energy because it corresponds to the Gibbs free energy when dealing with dilute solutions.
If we considered gas phases in a constant volume, $\freeenergy$ would be the Helmholtz free energy.
For instance, Debye-H\"uckel theory~\cite{Mcquarrie2000} derives $\freeenergy(\conc, \parameter)$ for electrolyte solutions, and van der Waals theory~\cite{Prigogine2015} provides $\freeenergy(\conc, \parameter)$ for a gas of interacting and finite-size molecules.
Standard computational approaches are also available to compute the free energy of generic non-ideal systems~\cite{Chipot2007,Lelievre2010}.
Without loss of generality, $\freeenergy(\conc,\parameter)$ can be written as the sum of an ideal term $\idfreeenergy(\conc)$ and a term accounting for the effects of the interactions $\nidfreeenergy(\conc,\parameter)$:
\begin{equation}
\freeenergy(\conc,\parameter) = \idfreeenergy(\conc) + \nidfreeenergy(\conc,\parameter)\,.
\label{eq_free_energy_zero_order}
\end{equation}

The free energy contribution carried by each species $\chemspecies$ is given by the chemical potential~\cite{holyst2012}:
\begin{equation}
\chempotential_{\chemspecies}(\conc) 
= \frac{\partial\freeenergy(\conc,\parameter)}{\partial[\chemspecies]}
=\idchempotential_\chemspecies([\chemspecies])+RT\ln\activitycoeff_\chemspecies(\conc,\parameter)\,.
\label{eq_elch_pot_0}
\end{equation}
By assumption, $\idchempotential_\chemspecies([\chemspecies])$ is the ideal chemical potential in either gas phases at constant volume $\volume$ or in dilute solutions,
\begin{equation}
\idchempotential_{\chemspecies}([\chemspecies])
= \frac{\partial\idfreeenergy(\conc)}{\partial[\chemspecies]}
=\stchempotential_{\chemspecies}+RT\ln[\chemspecies]\,,
\end{equation}
with $R$ the gas constant and $\stchempotential_{\chemspecies}$ the standard chemical potential.
The so-called activity coefficient $\activitycoeff_\chemspecies(\conc,\parameter)$ accounts for the effects of the interactions
\begin{equation}
RT\ln \activitycoeff_\chemspecies(\conc,\parameter) = \frac{\partial \potenergy(\conc,\parameter)}{\partial[\chemspecies]}\,.
\label{eq_activity_coeff}
\end{equation}
In equilibrium thermodynamics of non-ideal CRNs, $\activitycoeff_\chemspecies(\conc,\parameter)$ is often treated as a constant parameter, while in our framework it evolves in time since the concentrations $\conc$ are dynamical variables. 
The specific expression of $\activitycoeff_\chemspecies(\conc,\parameter)$, or equivalently of $\potenergy(\conc,\parameter)$, depends on the particular model used to describe the interactions.
We develop our theory only assuming that $\freeenergy(\conc,\parameter)$ is lower bounded and increases superlinearly as concentrations go to infinity.
These constraints are necessary to guarantee thermodynamic consistency, i.e., detailed balanced CRNs relax towards an equilibrium state, as we show in Subs.~\ref{sub_bound}. 
The free energy provided by the Debye-H\"uckel theory~\cite{Mcquarrie2000} satisfies these conditions: as concentrations become infinitely large, $\potenergy(\conc,\parameter)$ goes to minus infinity slower than $\idfreeenergy(\conc)$ goes to infinity.

We collect all the chemical potentials in the vector
\begin{equation}
\boldsymbol \chempotential (\conc) = \big(\dots,\chempotential_{\chemspecies}(\conc),\dots\big)^\intercal\,,
\end{equation}
that can be written as
\begin{equation}
\begin{split}
\boldsymbol \chempotential (\conc) 
&=\boldsymbol \stchempotential+RT\ln\conc+RT \ln \activitycoeffvec(\conc,\parameter)\\
& =\boldsymbol \stchempotential+RT\ln\conc+\boldsymbol\nabla_\conc\potenergy(\conc, \parameter)\,,
\end{split}
\label{eq_elch_pot}
\end{equation}
with $\boldsymbol \stchempotential=(\dots,\stchempotential_{\chemspecies},\dots)^\intercal$, $\activitycoeffvec(\conc,\parameter) = (\dots,\activitycoeff_\chemspecies(\conc,\parameter) ,\dots)^\intercal$ and $\boldsymbol\nabla_\conc =(\dots,\partial/\partial[\chemspecies],\dots)^\intercal$.

\begin{example}
We assume that the only relevant interactions in the CRN~\eqref{eq_crn_example} are the electrostatic interactions between charged species inside the same compartment.
For illustrative reasons, we describe the electrostatic potential in each compartment as a linear function of the net charge density instead of using the Debye-H\"uckel theory~\cite{Mcquarrie2000}.
This is consistent with a standard model of the transmembrane potential in mitochondria~\cite{Magnus1997,Magnus1998,Fall2001}.
The interaction free energy $\nidfreeenergy(\conc,\parameter)$ is thus given by the electrostatic potential energy.
It can be written as
\begin{equation}
\potenergy (\conc,\parameter) = \potenergy_{\compone}(\conc,\parameter) + \potenergy_\comptwo(\conc,\parameter)\,,
\label{eq_crn_example_int_pot_energ_tot}
\end{equation}
where  $\potenergy_{(\muteindex)}(\conc,\parameter)$, i.e., the electrostatic potential energy in the $\muteindex$ compartment, is a quadratic function of the net charge density in the $\muteindex$ compartment:
\begin{equation}
\potenergy_{(\muteindex)}(\conc,\parameter) = \frac{F\constant}{2}\big(\underbrace{-[\ch{A^{-}_{(\muteindex)}}]+[\ch{H^{+}_{(\muteindex)}}]+[\ch{C^{+}_{(\muteindex)}}]-[\ch{D^{-}_{(\muteindex)}}]}_{\equiv\chargedensity_{(\muteindex)}(\conc,\parameter)}\big)^2\,,
\label{eq_crn_example_int_pot_energ}
\end{equation}
with $F$ the Faraday constant and $\constant$ a generic proportionality constant.
On the membrane (compartment~$\compthr$), $\potenergy_\compthr=0$ since the chemical species are not charged and do not interact. 
Here, the parameters $\parameter$ are the concentrations of the species that are not involved in the chemical reactions, i.e., $\parameter = \big([\ch{C^{+}_\compone}],[\ch{C^{+}_\comptwo}],[\ch{D^{-}_\compone}],[\ch{D^{-}_\comptwo}]\big)$, but interact with species $\boldsymbol \chemspecies$. 
These concentrations may be controlled with some external processes.

 The chemical potentials of the species in the CRN~\eqref{eq_crn_example} are specified as
\begin{equation}
\begin{split}
&\chempotential_{\ch{E_\compthr}}(\conc) = \idchempotential_{\ch{E_\compthr}}([\ch{E_\compthr}]) = \stchempotential_{\ch{E}}+RT\ln[\ch{E_\compthr}]\,, \\
&\chempotential_{\ch{ES_\compthr}}(\conc) = \idchempotential_{\ch{ES_\compthr}}([\ch{ES_\compthr}]) = \stchempotential_{\ch{ES}}+RT\ln[\ch{ES_\compthr}]\,,\\
&\chempotential_{\ch{EP_\compthr}}(\conc) = \idchempotential_{\ch{EP_\compthr}}([\ch{EP_\compthr}]) = \stchempotential_{\ch{EP}}+RT\ln[\ch{EP_\compthr}]\,,\\
&\chempotential_{\ch{A^{-}_\compone}}(\conc) = \stchempotential_{\ch{A^{-}}}+RT\ln[\ch{A^{-}_\compone}] - F\constant \chargedensity_\compone(\conc,\parameter)\,,\\
&\chempotential_{\ch{A^{-}_\comptwo}}(\conc) = \stchempotential_{\ch{A^{-}}}+RT\ln[\ch{A^{-}_\comptwo}] - F\constant \chargedensity_\comptwo(\conc,\parameter)\,,\\
&\chempotential_{\ch{H^{+}_\compone}}(\conc) = \stchempotential_{\ch{H^{+}}}+RT\ln[\ch{H^{+}_\compone}] + F\constant \chargedensity_\compone(\conc,\parameter)\,,\\
&\chempotential_{\ch{H^{+}_\comptwo}}(\conc) = \stchempotential_{\ch{H^{+}}}+RT\ln[\ch{H^{+}_\comptwo}] + F\constant \chargedensity_\comptwo(\conc,\parameter)\,,\\
&\chempotential_{\ch{S_\comptwo}}(\conc) = \idchempotential_{\ch{S_\comptwo}}([\ch{S_\comptwo}]) = \stchempotential_{\ch{S}}+RT\ln[\ch{S_\comptwo}]\,,\\
&\chempotential_{\ch{P_\comptwo}}(\conc) = \idchempotential_{\ch{P_\comptwo}}([\ch{P_\comptwo}]) = \stchempotential_{\ch{P}}+RT\ln[\ch{P_\comptwo}]\,,
\end{split}
\end{equation}
where we assumed that the standard chemical potentials do not depend on the compartment, and the net charge densities $\big\{\chargedensity_{(\muteindex)}(\conc,\parameter)\big\}$ are specified in Eq.~\eqref{eq_crn_example_int_pot_energ}.
Note that the chemical potentials of the species $\ch{E_\compthr}$, $\ch{ES_\compthr}$, $\ch{EP_\compthr}$, $\ch{S_\comptwo}$ and $\ch{P_\comptwo}$ are the same as for ideal CRNs.
This is a consequence of accounting only for the interactions between charge species.
\qed
\end{example}

\subsection{Local Detailed Balance\label{sub_ldb}}
Here, we build the connection between dynamics and nonequilibrium thermodynamics. 
We use the \textit{local detailed balance} condition that binds the ratios between the forward and backward currents $\{\curr^{+\elrct}(\conc)/\curr^{-\elrct}(\conc)\}$ to the free energies of reaction $\{\freerct(\conc)\}$:
\begin{equation}
RT\ln\frac{\curr^{+\elrct}(\conc)}{\curr^{-\elrct}(\conc)}=-\freerct (\conc)\,.
\label{eq_local_detailed_balance_thermo}
\end{equation}
The latter is given by
\begin{equation}
\freerct (\conc)= \boldsymbol\chempotential(\conc)\cdot\matS_\elrct\,.
\label{eq_free_energy_reaction}
\end{equation}
This, together with Eq.~\eqref{eq_no_mass_action}, implies a constraint for the kinetic constants $\nik_{\pm\elrct}(\conc)$ of the chemical reactions:
\begin{equation}
\begin{split}
RT\ln\frac{\nik_{+\elrct}(\conc)}{\nik_{-\elrct}(\conc)}
&=-(\boldsymbol\stchempotential +RT\ln\activitycoeffvec(\conc,\parameter) )\cdot{\matS_\elrct} \\
&=-(\boldsymbol\stchempotential+\boldsymbol\nabla_\conc\potenergy(\conc, \parameter))\cdot\matS_\elrct\,.
\end{split}
\label{eq_local_detailed_balance_thermo_kinetic_constant}
\end{equation}
Equation~\eqref{eq_local_detailed_balance_thermo_kinetic_constant} shows that the $\conc$ dependence of the kinetic constants, which represents the breakdown of mass-action kinetics for non-ideal CRNs, comes from the interactions.
These can affect the relative stability of reagents and products and, consequently, the relative rate of the forward and backward reactions.
However, the local detailed balance condition highlights only the $\conc$ dependence of the antisymmetric part $\sqrt{\nik_{+\elrct}(\conc)/\nik_{-\elrct}(\conc)}$ of the kinetic constants.
This does not exclude that also the symmetric part $\sqrt{\nik_{+\elrct}(\conc)\nik_{-\elrct}(\conc)}$ of the kinetic constants is $\conc$ dependent as already discussed in the literature~\cite{Butt1999}.
In the limit $\activitycoeffvec(\conc,\parameter)\to1$, or equivalently $\boldsymbol\nabla_\conc\potenergy(\conc, \parameter)\to 0$, the local detailed balance condition for ideal CRNs is recovered,
\begin{equation}
RT\ln\frac{\ik_{+\elrct}}{\ik_{-\elrct}}=-\boldsymbol\stchempotential\cdot\matS_\elrct \,,
\end{equation}
and, consequently, the dynamics satisfies mass-action kinetics~\eqref{eq_mass_action}.

\begin{example}
We specify Eq.~\eqref{eq_local_detailed_balance_thermo_kinetic_constant} for the CRN~\eqref{eq_crn_example} assuming that the standard chemical potentials of~\ch{A^{-}} and~\ch{H^{+}} do not depend on the compartment (namely $\stchempotential_{\ch{A^{-}_{\compone}}}=\stchempotential_{\ch{A^{-}_{\comptwo}}}$ and $\stchempotential_{\ch{H^{+}_{\compone}}}=\stchempotential_{\ch{H^{+}_{\comptwo}}}$):
{\small
\begin{align}
&RT\ln\frac{\nik_{+1}(\conc)}{\nik_{-1}(\conc)}=F\constant\big(\chargedensity_\comptwo(\conc,\parameter)-\chargedensity_\compone(\conc,\parameter)\big)\,,\\
&RT\ln\frac{\nik_{+2}(\conc)}{\nik_{-2}(\conc)}=RT\ln\frac{\ik_{+2}}{\ik_{-2}}=-(\stchempotential_{\ch{ES_{\compthr}}}-\stchempotential_{\ch{E_{\compthr}}}-\stchempotential_{\ch{S_{\comptwo}}})\,,\\
&RT\ln\frac{\nik_{+3}(\conc)}{\nik_{-3}(\conc)}=-\big(\stchempotential_{\ch{EP_{\compthr}}}-\stchempotential_{\ch{ES_{\compthr}}}+F\constant\big(\chargedensity_\comptwo(\conc,\parameter)-\chargedensity_\compone(\conc,\parameter)\big)\big)\,,\\
&RT\ln\frac{\nik_{+4}(\conc)}{\nik_{-4}(\conc)}=RT\ln\frac{\ik_{+4}}{\ik_{-4}}=-(\stchempotential_{\ch{P_{\comptwo}}}+\stchempotential_{\ch{E_{\compthr}}}-\stchempotential_{\ch{EP_{\compthr}}})\,.
\end{align}}
\qed
\end{example}

\subsection{Chemostatting\label{sub_chemostat}}
Thermodynamically, the chemostatting procedure directly determines the chemical potentials $\{\chempotential_\chemspecies\}$ of the $\setchemostatted$ species.
For ideal CRNs, where $\chempotential_\chemspecies(\conc)= \idchempotential_\chemspecies([\chemspecies])$, this means that the chemostatting procedure directly determines the concentrations, $[\chemspecies]=\exp((\chempotential_\chemspecies-\stchempotential_\chemspecies)/RT)$.
For non-ideal CRNs, the equivalence 
\begin{equation}
\chempotential_\chemspecies =  \stchempotential_{\chemspecies}+RT\ln[\chemspecies]+RT\ln\activitycoeff_\chemspecies(\conc,\parameter)
\text{ }\text{ }\text{ }\forall\chemspecies\in\setchemostatted\,,
\label{eq_constaint_chemostats}
\end{equation}
does not directly determine the concentrations $\boldsymbol\exconc$ due to the dependence on the interactions, i.e.,
$[\chemspecies]=\exp((\chempotential_\chemspecies-\stchempotential_\chemspecies)/RT)/\activitycoeff_\chemspecies(\conc,\parameter)$ 
$\forall\chemspecies\in \setchemostatted$.
Hence, $\boldsymbol \excurr(t)$ in Eq.~\eqref{eq_dynamics_crns} (or $\boldsymbol \excurrY(t)$ in Eq.~\eqref{eq_dynamics_crns_Y}) describes the exchange currents such that the chemical potentials of the $\setchemostatted$ species are externally determined. 

We now derive the explicit form of $\boldsymbol\excurrY(t)$.
The readers who may want to skip the details of the derivation can go directly to Eq.~\eqref{eq_final_I_expression}.
The constraints imposed by Eq.~\eqref{eq_constaint_chemostats} must be satisfied for every time $t$ of the dynamics.
Thus, also their time derivative must vanish:
\begin{equation}
\begin{split}
\frac{RT}{[\chemspecies](t)}\dt[\chemspecies](t)+\boldsymbol\nabla_\conc \potential_{\chemspecies}\big(\conc(t), \parameter(t)\big)\cdot\dt\conc(t) +&\\
+\boldsymbol\nabla_\parameter \potential_{\chemspecies}\big(\conc(t), \parameter(t)\big)\cdot\dt \parameter(t)- \dt\chempotential_{\chemspecies}(t)&=0\,.
\end{split}
\label{eq_derivation_excurr_1} 
\end{equation}
Here, we used the following compact notation $\potential_{\chemspecies}(\conc(t), \parameter(t))=RT\ln\activitycoeff_\chemspecies(\conc(t),\parameter(t))$ and we accounted for the possible time-dependent control of the chemical potentials $\{\chempotential_{\chemspecies}\}$ and of the parameters $\parameter$.
The set of Eqs.~\eqref{eq_derivation_excurr_1} for every $\chemspecies\in\setchemostatted$ can be rewritten as 
\begin{equation}
\matW(t)(\matS \boldsymbol\curr(\conc(t))+\boldsymbol\excurr(t)) + \boldsymbol \fluxexmolarpotenergy (t)- \dt\boldsymbol\chempotential_{\setchemostatted}(t) =0\,,
\end{equation}
by introducing the vectors 
\begin{equation}
\boldsymbol \fluxexmolarpotenergy (t)= \left(\dots, \boldsymbol\nabla_\parameter \potential_{\chemspecies}\big(\conc(t), \parameter(t)\big)\cdot \dt\parameter(t),\dots\right)^\intercal_{\chemspecies\in\setchemostatted}
\end{equation}
and $\boldsymbol\chempotential_{\setchemostatted}(t)=(\dots,\chempotential_{\chemspecies}(t) ,\dots)^\intercal_{\chemspecies\in\setchemostatted}$, and the matrix $\matW(t)$ whose entries are
\begin{equation}
\matW^\chemspecies_\chemspeciesB (t)= \frac{RT}{[\chemspecies](t)}\delta_{\chemspecies, \chemspeciesB} + \frac{\partial\potential_{\chemspecies}\big(\conc(t), \parameter(t)\big)}{\partial[\chemspeciesB]}\,,
\label{eq_matrix_W_current}
\end{equation}
with $\chemspecies \in \setchemostatted$ the row index and $\chemspeciesB \in \setchemspecies$ the column index. We notice that the submatrix $\matWY$, whose entries are $\{\matW^\chemspecies_\chemspeciesB\}_{\chemspecies, \chemspeciesB\in \setchemostatted}$, is square and not nonsingular so that it can be inverted. We thus obtain an explicit expression for $\boldsymbol\excurrY(t)$:
\begin{equation}
\boldsymbol\excurrY(t) =- \Big(\matWY(t)\Big)^{-1}\Big(\matW(t)\matS \boldsymbol\curr(\conc(t)) + \boldsymbol \fluxexmolarpotenergy (t)- \dt\boldsymbol\chempotential_{\setchemostatted}(t) \Big)\,.
\label{eq_final_I_expression}
\end{equation}
Systems are said to be \textit{autonomous} when the chemostatting procedure maintains the chemical potentials $\boldsymbol\chempotential_{\setchemostatted}$ and the parameters $\parameter$ constant, i.e., $\dt\boldsymbol\chempotential_{\setchemostatted}(t) = 0$ and $\boldsymbol \fluxexmolarpotenergy (t) =0$.
Otherwise, systems are said to be \textit{nonautonomous}.
The matrices $\matW(t)$ and $\matWY(t)$ link the value of the external currents to the concentrations of the internal species that interact with the chemostatted ones. 
In the limit $\activitycoeffvec(\conc,\parameter)\to1$, or equivalently $\boldsymbol\nabla_\conc\potenergy(\conc, \parameter)\to 0$,  $\boldsymbol\excurrY(t) $ for ideal CRNs is recovered
\begin{equation}
\begin{split}
\boldsymbol\excurrY(t) &=-\matSY \boldsymbol\curr(\conc(t)) + \Big(\matWY(t)\Big)^{-1} \dt  \idchempotentialY(t) \\
& =-\matSY \boldsymbol\curr(\conc(t)) + \dt  \boldsymbol\exconc(t) \,,
\end{split}
\end{equation}
controlling directly the concentrations~\cite{Rao2016} according to the protocol $\dt  \boldsymbol\exconc(t)$.

\begin{example}
We specify the different terms in Eq.~\eqref{eq_final_I_expression} for the CRN~\eqref{eq_crn_example}.
According to Eq.~\eqref{eq_matrix_W_current}, $\matW(t)$ is given by
\begin{equation}
\scaleto{
\matW = 
 \kbordermatrix{
    & \color{g}\ch{E_{\compthr}} &\color{g}\ch{ES_{\compthr} }&\color{g}\ch{EP_{\compthr} }&\color{g}\ch{A^{-}_{\compone}}& \color{g}\ch{A^{-}_{\comptwo}}&&\color{g}\ch{H^{+}_{\compone}}&\color{g}\ch{H^{+}_{\comptwo}}&\color{g}\ch{S_{\comptwo}}&\color{g}\ch{P_{\comptwo}}\cr
    \color{g}\ch{H^{+}_{\compone}} & 0 &0& 0 & {-F\constant}&0 &\color{g}\color{g}\vrule&{\frac{RT}{[\ch{H^{+}_{{\compone}}}]} + F\constant}&0 & 0&0\cr
    \color{g}\ch{H^{+}_{\comptwo}} &0 &0& 0 & 0&{-F\constant}&\color{g}\color{g}\vrule &0&{\frac{RT}{[\ch{H^{+}_{{\comptwo}}}]} + F\constant} & 0&0\cr
    \color{g}\ch{S_{\comptwo}}         &0 &0& 0 & 0&0 &\color{g}\color{g}\vrule&0& 0& {\frac{RT}{[\ch{S_{{\comptwo}}}]}} &0\cr
    \color{g}\ch{P_{\comptwo}}         &0 &0& 0 & 0&0 &\color{g}\color{g}\vrule&0&0 & 0&{\frac{RT}{[\ch{P_{{\comptwo}}}]}}\cr
  }}{60pt}\,,
\label{eq_crn_example_matW}
\end{equation}
and hence $(\matWY(t))^{-1}$ is given by 
\begin{equation}
(\matWY)^{-1} = 
 \kbordermatrix{
    &\color{g}\ch{H^{+}_{\compone}}&\color{g}\ch{H^{+}_{\comptwo}}&\color{g}\ch{S_{\comptwo}}&\color{g}\ch{P_{\comptwo}}\cr
    \color{g}\ch{H^{+}_{\compone}} &{\frac{[\ch{H^{+}_{{\compone}}}]}{RT + F \constant[\ch{H^{+}_{{\compone}}}]}}&0 & 0&0\cr
    \color{g}\ch{H^{+}_{\comptwo}} &0&{\frac{[\ch{H^{+}_{{\comptwo}}}]}{RT + F \constant[\ch{H^{+}_{{\comptwo}}}]}}& 0&0\cr
    \color{g}\ch{S_{\comptwo}}         &0& 0& {\frac{[\ch{S_{{\comptwo}}}]}{RT}} &0\cr
    \color{g}\ch{P_{\comptwo}}         &0&0 & 0&{\frac{[\ch{P_{{\comptwo}}}]}{RT}}\cr
}
\,.
\label{eq_crn_example_inv_matWY}
\end{equation}
Here, we did not write the explicit time dependence of the concentrations and of the matrices for the sake of compactness.
The vector $\boldsymbol \fluxexmolarpotenergy (t)$ is specified as
\begin{equation}
\boldsymbol \fluxexmolarpotenergy (t)=
 \kbordermatrix{
     & \cr
    \color{g}\ch{H^{+}_{\compone}}   &F\constant \big(\dt[\ch{C^+_\compone}](t)-\dt[\ch{D^-_\compone}](t)\big)\cr
    \color{g}\ch{H^{+}_{\comptwo}} &F\constant  \big(\dt[\ch{C^+_\comptwo}](t)-\dt[\ch{D^-_\compone}](t)\big)\cr
    \color{g}\ch{S_{\comptwo}}   &0\cr
    \color{g}\ch{P_{\comptwo}}   &0\cr
  }\,.
\end{equation}
\qed
\end{example}

\subsection{Decomposition of the Entropy Production Rate\label{sub_epr_decomposition}}
We split the entropy production rate into different contributions.
To do so we adapt the decomposition of the entropy production for stochastic ideal CRNs developed in Ref.~\onlinecite{Rao2018b} to deterministic non-ideal CRNs. 
This is not straightforward  and represents a major result of this work.
The readers who may want to skip the details of the derivation can go directly to Eq.~\eqref{eq_epr_decomposition}.

We start by expressing the entropy production rate of deterministic CRNs~\cite{Polettini2014, Rao2016},
\begin{equation}
\epr(t) = R \sum_{\elrct\in\setelrct}\left(\curr^{+\elrct}(\conc(t))-\curr^{-\elrct}(\conc(t))\right)\ln\frac{\curr^{+\elrct}(\conc(t))}{\curr^{-\elrct}(\conc(t))}\geq 0\,.
\end{equation}
Using the local detailed balance~\eqref{eq_local_detailed_balance_thermo} and the specific expressions for the free energies of reaction~\eqref{eq_free_energy_reaction}, we get
\begin{equation}
T\epr (t)= - \boldsymbol\chempotential(\conc(t))\cdot\matS\boldsymbol\curr(\conc(t))\,.
\label{eq_epr_thermo}
\end{equation}
By explicitly considering the contributions of the $\setinternal$, $\setforce$ and $\setpotential$ species, we obtain
\begin{equation}
\small
T\epr (t)= - \left(\boldsymbol\chempotential_\setinternal(\conc(t))\cdot\matSX + \boldsymbol\chempotential_\setforce(t)\cdot\matSYf+ \boldsymbol\chempotential_\setpotential(t)\cdot\matSYp\right)\boldsymbol\curr(\conc(t))\,,
\label{eq_initial_dec_epr}
\end{equation}
with $\boldsymbol\chempotential_{\setvariable}(\bullet)=(\dots, \chempotential_\chemspecies(\bullet),\dots)^\intercal_{\chemspecies\in\setvariable}$ and $\setvariable=\{\setinternal, \setforce, \setpotential\}$. 
Note that we do not explicitly write the $\conc(t)$ dependence of the chemical potentials of the $\setchemostatted$ species since they are externally controlled.

A part of the free energy externally exchanged through the currents $\{\excurr^\chemspecies(t)\}$ modifies the free energy of the system.
Another part is transferred through the system.
To proceed with the entropy production decomposition, we need to distinguish these two parts.
To do so we exploit the conservation laws.
We consider
\begin{equation}
\matLb\matS = 0 = \matLbX\matSX + \matLbYf\matSYf + \matLbYp\matSYp\,,   
\end{equation}
and, recalling that the matrix $\matLbYp$ can be inverted (see Subs.~\ref{sub_cons_law}), we write
\begin{equation}
\matSYp = -(\matLbYp)^{-1}( \matLbX\matSX + \matLbYf\matSYf)\,.
\label{eq_link_yp_yf}
\end{equation}
Equation~\eqref{eq_link_yp_yf} binds the net variation of the number of molecules for the $\setpotential$ species to that of the $\setinternal$ and $\setforce$ species.
We substitute this result in Eq.~\eqref{eq_initial_dec_epr}:
\begin{equation}
\begin{split}
T\epr (t)= &- \Big(\boldsymbol\chempotential_\setinternal(\conc(t))\cdot\matSX + \boldsymbol\chempotential_\setforce(t)\cdot\matSYf+\\
&-\boldsymbol\chempotential_\setpotential(t)\cdot(\matLbYp)^{-1}( \matLbX\matSX + \matLbYf\matSYf)    \Big)\boldsymbol\curr(\conc(t))\,.
\end{split}
\end{equation}
By adding and subtracting $\boldsymbol\chempotential_\setpotential(t)\cdot(\matLbYp)^{-1}\matLbYp\matSYp$, the entropy production rate becomes
\begin{equation}
T\epr (t)= - \molarstatefun(\conc(t) )\cdot\matS\boldsymbol\curr(\conc(t))\,,
\label{eq_inter1_dec_epr}
\end{equation}
where
\begin{equation}
\molarstatefun(\conc(t) )=  \big(\boldsymbol\chempotential(\conc(t))\cdot\one - \boldsymbol\chempotential_\setpotential(t)\cdot(\matLbYp)^{-1}\matLb \big)^\intercal
\end{equation}
with $\one$ the identity matrix. Note that we just introduced a term in the expression of the entropy production rate, i.e., $\boldsymbol\chempotential_\setpotential(t)\cdot(\matLbYp)^{-1}\matLb$, whose net contribution vanishes since $\matLb \matS =0$. Introducing this term allows us to recognize the proper thermodynamic potential of open non-ideal CRNs.

To do so, we define the following state function
\begin{equation}
\statefun(\conc(t))= \molarstatefun(\conc(t))\cdot\conc(t)\,,
\end{equation}
whose time derivative according to the rate equation Eq.~\eqref{eq_dynamics_crns} reads
\begin{equation}
\begin{split}
\dt\statefun(\conc(t))&=  \molarstatefun(\conc(t))\cdot\matS \boldsymbol \curr(\conc(t))  + \molarstatefunY(t)\cdot\boldsymbol \excurrY(t) +\\
& + RT \dt\norm{\conc(t)} - \dt\boldsymbol\chempotential_\setpotential(t)\cdot(\matLbYp)^{-1}\matLb\conc(t)+\\
&+ \dt\big(\boldsymbol\nabla_{\conc} \potenergy (\conc(t),\parameter(t))\big)\cdot\conc(t)\,,
\end{split}
\label{eq_inter2_dec_epr}
\end{equation}
where $\molarstatefunY(t) = (\dots, \molarstatefunscalar_\chemspecies(t),\dots)_{\chemspecies\in\setchemostatted}$, $\norm{\conc(t)} = \sum_{\chemspecies}[\chemspecies](t)$.
The time dependence of $\parameter(t)$ accounts for the (possible) time dependent manipulation of the parameters.

To procede with the decomposition, we first recognize that the term $ \molarstatefunY(t)\cdot\boldsymbol \excurrY(t)$ simplifies to $ \molarstatefunYf(t)\cdot\boldsymbol \excurrYf(t)$ since $\molarstatefunscalar_\chemspecies=0$ for $\chemspecies \in \setpotential$.
We then notice that the term $\dt\big(\boldsymbol\nabla_{\conc} \potenergy (\conc(t),\parameter(t))\big)\cdot\conc(t)$ can be written as
\begin{align}
\dt\big(&\boldsymbol\nabla_{\conc} \potenergy (\conc(t),\parameter(t))\big)\cdot\conc(t)= \nonumber\\
 =& \dt\big(\boldsymbol\nabla_{\conc} \potenergy (\conc(t),\parameter(t))\cdot\conc(t)\big) - \boldsymbol\nabla_{\conc} \potenergy (\conc(t),\parameter(t))\cdot\dt\conc(t)\nonumber \\
=& \dt\big(\boldsymbol\nabla_{\conc} \potenergy (\conc(t),\parameter(t))\cdot\conc(t)\big) - \dt \potenergy (\conc(t),\parameter(t))+\nonumber\\
&+\boldsymbol\nabla_{\parameter} \potenergy (\conc(t),\parameter(t))\cdot\dt\parameter(t)
\end{align}
by using the time derivative of the interaction free energy
\begin{equation}
\begin{split}
\dt\potenergy(\conc(t),\parameter(t))= &\boldsymbol\nabla_{\conc} \potenergy (\conc(t),\parameter(t))\cdot\dt\conc(t)+\\
&+\boldsymbol\nabla_{\parameter} \potenergy (\conc(t),\parameter(t))\cdot\dt\parameter(t)\,,
\end{split}
\end{equation}
with $\boldsymbol\nabla_{\parameter} = (\dots,\partial/\partial\scalarparameter_\muteindex,\dots)^\intercal$.

We now exploit Eq.~\eqref{eq_inter2_dec_epr} to express $ \molarstatefun(\conc(t))\cdot\matS \boldsymbol \curr(\conc(t))$ as a function of the other terms and we substitute it in Eq.~\eqref{eq_inter1_dec_epr}. By doing so and collecting all the full time derivatives in a single term denoted $\dt\semigrand$, we obtain the following decomposition of the entropy production rate, which constitutes a central result of this work,
\begin{equation}
T\epr(t)=-\dt \semigrand(\conc(t)) + \ncwr(t) + \dw(t)\geq0\,.
\label{eq_epr_decomposition}
\end{equation}

Here, we introduced the semigrand free energy of non-ideal systems as
\begin{equation}
\begin{split}
\semigrand (\conc)=& \boldsymbol\chempotential(\conc)\cdot\conc 
- RT \norm{\conc}
- \boldsymbol\chempotential_\setpotential(t)\cdot\boldsymbol \moieties(\conc)+\\ 
&+ \potenergy(\conc,\parameter)
-\boldsymbol\nabla_{\conc} \potenergy(\conc, \parameter)\cdot\conc
\end{split}
\label{eq_semigrand}
\end{equation}
with the concentrations of the so-called moieties:
\begin{equation}
\boldsymbol \moieties(\conc) = (\matLbYp)^{-1}\matLb \conc\,.
\label{eq_moieties}
\end{equation} 
They represent parts of (or entire) molecules which are exchanged with the environment. 
The semigrand free energy~\eqref{eq_semigrand} can be rewritten in a more appealing ways as 
\begin{equation}
\begin{split}
\semigrand (\conc) 
&= \boldsymbol\idchempotential(\conc)\cdot\conc - RT \norm{\conc}- \boldsymbol\chempotential_\setpotential(t)\cdot\boldsymbol \moieties(\conc)+  \potenergy(\conc,\parameter)\\
&=\freeenergy(\conc, \parameter)- \boldsymbol\chempotential_\setpotential(t)\cdot\boldsymbol \moieties(\conc)
\end{split}
\label{eq_semi_grand_2}
\end{equation}
by using Eq.~\eqref{eq_elch_pot}. The term $\boldsymbol\idchempotential(\conc)\cdot\conc - RT \norm{\conc}$ is the free energy of closed ideal CRNs, i.e., $\idfreeenergy(\conc)$, accounting for the chemical energy of every species. 
The term $\potenergy(\conc,\parameter)$ is the interaction free energy. 
Their sum gives the free energy $\freeenergy(\conc,\parameter)$ in Eq.~\eqref{eq_free_energy_zero_order}.  
The term $\boldsymbol\chempotential_\setpotential(t)\cdot\boldsymbol \moieties(\conc)$ is  the energetic contribution of the matter exchanged with the environment. 
It vanishes for closed systems.
Since $\semigrand (\conc) $ is a state function, its time derivative vanishes at steady-state.

The driving work rate $\dw(t)$ can be split into the sum of the chemical and the interaction driving work rates $\dw(t) = \cdw(t) + \edw(t)$ that are are specified as 
\begin{equation}
\cdw(t) = - \dt\boldsymbol\chempotential_\setpotential(t)\cdot \boldsymbol  \moieties(\conc(t))
\label{eq_ch_driving}
\end{equation}
and
\begin{equation}
\edw(t) =  \boldsymbol\nabla_{\parameter} \potenergy (\conc(t),\parameter(t))\cdot\dt\parameter(t)\,,
\label{eq_el_driving}
\end{equation}
respectively.
The chemical driving work~\eqref{eq_ch_driving} accounts for the time dependent control of only the chemical potentials $\boldsymbol\chempotential_\setpotential(t)$. 
The interaction driving work~\eqref{eq_el_driving} accounts for the time dependent control of the interaction free energy through the time evolution of the parameters $\parameter(t)$.
They represent the energetic cost of modifying the equilibrium that is defined if  $\setchemostatted = \setpotential$ (as shown in Subs.~\ref{sub_equilibrium}, open CRNs with $\setchemostatted = \setpotential$ have a well defined equilibrium state for every value of $\parameter$). 
These two terms vanish in autonomous systems.

The nonconservative work rate is given by 
\begin{equation}
\ncwr(t) = \ncforce(t) \cdot\boldsymbol \excurrYf(t)\,.
\label{eq_nc_work}
\end{equation}
It quantifies the energetic cost to sustain fluxes of chemical species through the CRN by means of the fundamental nonconservative forces 
\begin{equation}
\ncforce(t) = \left(\boldsymbol\chempotential_\setforce(t)\cdot\one - \boldsymbol\chempotential_\setpotential(t)\cdot(\matLbYp)^{-1}\matLbYf \right)^\intercal\,.
\label{eq_nc_force}
\end{equation}
These are the forces keeping the system out of equilibrium. 
Indeed, they vanish when the CRN is detailed balanced.
This always happens when all the chemostatted species break a conservation law $\setchemostatted = \setpotential$ and, for this reason, the CRNs are said \textit{unconditionally} detailed balanced.
Therefore, the $\setforce$ species are named \textit{force} species.
The $\setpotential$ species are named \textit{potential} species since they fix the equilibrium conditions.
Thus, the splitting $\setchemostatted = \setforce \cup\setpotential$ must be motivated by the different thermodynamic role of the $\setforce$ species and $\setpotential$ species.
The nonconservative work accounts for autonomous mechanisms keeping the system out of equilibrium unlike the chemical driving work~\eqref{eq_ch_driving} and the interaction driving work~\eqref{eq_el_driving} .
We stress that the expression of the fundamental force~\eqref{eq_nc_force} for non-ideal CRNs is analogous to that found for ideal CRNs~\cite{Rao2018b}.
The effects of the interactions that modify the value of the nonconservative work~\eqref{eq_nc_work} are hidden in the expressions of the exchanged currents~\eqref{eq_final_I_expression}. 
Note also that the force~\eqref{eq_nc_force} can vanish, and hence the CRN becomes detailed balanced, even if $\setchemostatted \neq \setpotential$. 
Indeed, $\ncforce = 0$ when the chemical potentials of the $\setforce$ species satisfy 
$\boldsymbol\chempotential_\setforce^\intercal =  \boldsymbol\chempotential_\setpotential\cdot(\matLbYp)^{-1}\matLbYf$.
Thus, the fundamental nonconservative forces are generated by the difference between the actual values of chemical potentials of the $\setforce$ species, i.e., $\boldsymbol\chempotential_\setforce$,  and their equilibrium values if only the $\setpotential$ species were chemostatted, i.e., $\boldsymbol\chempotential_\setpotential\cdot(\matLbYp)^{-1}\matLbYf$.

Finally, we notice that for autonomous ($\cdw(t)=\edw(t) = 0$) detailed balanced ($\ncwr(t)=0$) CRNs, the semigrand free energy decreases monotonously in time, i.e., $\dt\semigrand(t)=-T\epr(t)\leq0$. The semigrand free energy is dissipated. 

\begin{example}
We specify the moieties~\eqref{eq_moieties}, the interaction driving work~\eqref{eq_el_driving}, the nonconservative forces~\eqref{eq_nc_force} and  the nonconservative work~\eqref{eq_nc_work} for the open CRN~\eqref{eq_crn_example}. 
We start by giving an explicit expression for the matrix $(\matLbYp)^{-1}\matLb$ (with  $\matLb$ given in Eq.~\eqref{eq_crn_example_matL}):
\begin{equation}
(\matLbYp)^{-1}\matLb = 
 \kbordermatrix{
    & \color{g}\ch{E_{\compthr}} &\color{g}\ch{ES_{\compthr} }&\color{g}\ch{EP_{\compthr} }&\color{g}\ch{A^{-}_{\compone}}& \color{g}\ch{A^{-}_{\comptwo}}&&\color{g}\ch{H^{+}_{\compone}}&&\color{g}\ch{H^{+}_{\comptwo}}&\color{g}\ch{S_{\comptwo}}&\color{g}\ch{P_{\comptwo}}\cr
    \color{g}\ch{H_2^{+}} &0 &0& 0 & 0&0 &\color{g}\vrule&1& \color{g}\vrule&1 & 0&0\cr
    \color{g}\ch{S_2} &0 &1& 0 & 0&0&\color{g}\vrule &-1& \color{g}\vrule&0 & 1&0\cr
    \color{g}\ch{P_2}   &0 &0& 1 & 0&0 &\color{g}\vrule&1& \color{g}\vrule&0 & 0&1\cr
  }\,.
\label{eq_crn_example_matLbYpinv_matLb}
\end{equation}
The grey vertical lines mark the split of $(\matLbYp)^{-1}\matLb $ into $(\matLbYp)^{-1}\matLbX$, $(\matLbYp)^{-1}\matLbYf$, and $(\matLbYp)^{-1}\matLbYp=\one$.
By using Eq.~\eqref{eq_crn_example_matLbYpinv_matLb} in the definition of the concentrations of the moieties~\eqref{eq_moieties}, we obtain that
\begin{equation}
\boldsymbol \moieties(\conc) =
\begin{pmatrix}
[\ch{H^{+}_{\compone}}] + [\ch{H^{+}_{\comptwo}}]\\
[\ch{ES_{\compthr}}] + [\ch{S_{\comptwo}}] -[\ch{H^{+}_{\compone}}]\\
[\ch{EP_{\compthr}}] + [\ch{P_{\comptwo}}] +[\ch{H^{+}_{\compone}}]
\end{pmatrix}\,.
\end{equation}
Only the first moiety has a straightforward interpretation as the protons. 
The others are linear combination of broken conserved quantities accounting for the fragments of molecules exchange through the CRN.

We now consider the interaction driving work~\eqref{eq_el_driving}. 
By using the expression of the interaction free energy for this example~\eqref{eq_crn_example_int_pot_energ_tot}, we obtain
\begin{equation}\small
\begin{split}
\edw(t) = F\constant \Big(&\chargedensity_\compone(\conc(t),\parameter(t))\big(\dt[\ch{C^{+}_{\compone}}](t)-\dt[\ch{D^{-}_{\compone}}](t)\big)+\\
+&\chargedensity_\comptwo(\conc(t),\parameter(t))\big(\dt[\ch{C^{+}_{\comptwo}}](t)-\dt[\ch{D^{-}_{\comptwo}}](t)\big)\Big)\,,
\end{split}
\end{equation}
accounting for the variation of the semigrand free energy due to possible nonautonomous processes changing the concentrations of the cations $\ch{C^{+}}$ and the anions \ch{D^{-}} in the two compartments.

Finally, we turn to the nonconservative forces~\eqref{eq_nc_force}.
There is only one nonconservative force for the open CRN~\eqref{eq_crn_example},
\begin{equation}
\begin{split}
\ncforcescalar(t) &= \chempotential_{\ch{H^{+}_{\compone}}}(t) - \boldsymbol\chempotential_\setpotential(t)\cdot(\matLbYp)^{-1}\matLbYf \\
&= -(\idchempotential_{\ch{P_{\comptwo}}}(t)  + \chempotential_{\ch{H^{+}_{\comptwo}}}(t) - \chempotential_{\ch{H^{+}_{\compone}}}(t)-\idchempotential_{\ch{S_{\comptwo}}}(t))\,,
\end{split}
\label{eq_crn_example_ncforce}
\end{equation}
since $ \boldsymbol\chempotential_\setpotential(t)=(\chempotential_{\ch{H^{+}_{\comptwo}}}(t), \idchempotential_{\ch{S_{\comptwo}}}(t),\idchempotential_{\ch{P_{\comptwo}}}(t))^\intercal$ (recall that the chemical potential of $\ch{S_{\comptwo}}$ and $\ch{P_{\comptwo}}$ is the same as for ideal CRNs).
It represents the thermodynamic force (i.e., the free energy of reaction) of the following effective chemical reaction 
\begin{equation}
\ch{S_{\comptwo}  + H^{+}_{\compone} <=> H^{+}_{\comptwo} + P_{\comptwo} }
\label{eq_crn_example_eff_reaction_nonforce}
\end{equation}
between chemostatted species. 
Note that the role of the chemostatted species is precisely that of keeping this effective reaction out of equilibrium by imposing the thermodynamic force~\eqref{eq_crn_example_ncforce}.
The corresponding nonconservative work reads
\begin{equation}
\ncwr(t)=  -(\idchempotential_{\ch{P_{\comptwo}}}(t)  + \chempotential_{\ch{H^{+}_{\comptwo}}}(t) - \chempotential_{\ch{H^{+}_{\compone}}}(t)-\idchempotential_{\ch{S_{\comptwo}}}(t))\excurr^{\ch{H^{+}_{\compone}}}(t)\,.
\label{eq_crn_example_ncwr}
\end{equation}
We stress that, despite Eq.~\eqref{eq_crn_example_ncwr} resembling the corresponding expression for ideal CRNs, the interactions affect the value the exchange current~$\excurr^{\ch{H^{+}_{\compone}}}(t)$. 
\qed
\end{example}

\subsection{Steady-State Entropy Production~\label{sub_ss_entropy}}
When autonomous CRNs reach a nonequilibrium steady state $\steady{\conc}$, the entropy production rate~\eqref{eq_epr_decomposition} simplifies to 
\begin{equation}
T\steady{\epr}= \ncwrss =\ncforce\cdot\excurrYfss\,,
\label{eq_entropy_ss}
\end{equation}
highlighting the physical role of the nonconservative work: it provides the energy to balance dissipation and keep CRNs out of equilibrium.

An equivalent decomposition can be achieved using the cycle splitting of the current vector~\eqref{eq_ss_current} in Eq.~\eqref{eq_epr_thermo}:
\begin{equation}
\begin{split}
T\steady{\epr}
&=- \boldsymbol\chempotential\cdot\matS\steady{\boldsymbol\curr}\\
&=- \boldsymbol\chempotential\cdot\matS\bigg\{\sum_\internalindex\boldsymbol\cycle_\internalindex\coeffcycle^\internalindex+\sum_\emergentindex\boldsymbol\cycle_\emergentindex\coeffcycle^\emergentindex\bigg\}\\
&= - \boldsymbol\chempotential_\setchemostatted\cdot\effmatSY\boldsymbol \coeffcycle
\end{split}
\label{eq_epr_decomposition2}
\end{equation}
where we used Eq.~\eqref{eq_internal_cycles} and~\eqref{eq_emergent_cycles}.
The vector $\boldsymbol \coeffcycle$ collects the coefficients $\{\coeffcycle^\emergentindex\}$, i.e., $\boldsymbol \coeffcycle  = (\dots,\coeffcycle^\emergentindex,\dots)^{\intercal}$.
Each $\emergentindex$ column of the effective stoichiometric matrix $\effmatSY_\emergentindex =\matSY \boldsymbol\cycle_\emergentindex$ codifies the net stoichiometry of the chemostatted species along the $\emergentindex$ emergent cycle.

This alternative decomposition of the entropy production has two advantages.
First, it emphasizes the role of the emergent cycles in the steady-state dissipation.
They are out of equilibrium effective reactions interconverting chemostatted species according to the stoichiometry codified in $\effmatSY$.
Second, this decomposition can be used to develop thermodynamically consistent coarse-graining theories as it has been done for ideal CRNs~\cite{Wachtel2018, Avanzini2020b}.

\begin{example}
We specify the effective stoichiometric matrix $\effmatSY$ for the CRN~\eqref{eq_crn_example} given the emergent cycle~\eqref{eq_crn_example_emergent_cycle}:
\begin{equation}
{\effmatSY}=
 \kbordermatrix{
    & \cr
       \color{g}\ch{H^{+}_{\compone}}   &-1\cr
    \color{g}\ch{H^{+}_{\comptwo}} &1\cr
    \color{g}\ch{S_{\comptwo}}   &-1\cr
    \color{g}\ch{P_{\comptwo}}   &1\cr
  }\,.
  \label{eq_crn_example_effmatSY}
\end{equation}
Notice that the corresponding effective reaction is equal to the one we inferred from the nonconservative force in Eq.~\eqref{eq_crn_example_eff_reaction_nonforce}.
This equivalence does not hold in general.
\qed
\end{example}

\subsection{Lower Bound of the Semigrand Free Energy~\label{sub_bound}}
We now examine how the assumptions on the free energy $\freeenergy(\conc,\parameter)$ introduced in Subs.~\ref{sec_set_up_thermo} imply that the semigrand free energy  $\semigrand(\conc)$~\eqref{eq_semi_grand_2} is lower bounded and thus guarantee thermodynamic consistency.

Since $\freeenergy(\conc,\parameter)$ is lower bounded and increases superlinearly as concentrations go to infinity by assumption, also $\semigrand(\conc)$ (see Eq.~\eqref{eq_semi_grand_2}) is lower bounded for finite values of the chemical potential of the chemostatted species.
Indeed, as concentrations become infinitely large, $-\boldsymbol\chempotential_\setpotential\cdot\boldsymbol \moieties$ can go to minus infinity slower (linearly with $\conc$) than $\freeenergy(\conc,\parameter)$ goes to infinity (superlinearly with $\conc$).
Thus, $\exists \semigrand_\minl$ such that
\begin{equation}
\semigrand(\conc) \geq\semigrand_\minl\,,\text{ }\text{ }\text{ }\forall\conc\,.
\label{eq_g_lowest_bound}
\end{equation}
This ensures thermodynamic consistency: autonomous detailed balanced systems relax towards an equilibrium state.
Given that 
i) $\dt\semigrand(\conc(t)) = -T\epr(t)\leq0$ (see Subs.~\ref{sub_epr_decomposition}), 
ii) $\epr(t) = 0$ only at equilibrium and 
iii) $\semigrand(\conc(t))$ cannot decrease indefinitely because of Eq.~\eqref{eq_g_lowest_bound},
the system must reach an equilibrium state, $\lim_{t\to+\infty} \conc(t)=\conc_\eq$, and $\semigrand(\conc(t))\geq\semigrand(\conc_\eq)$, $\forall t$.

The specific equilibrium state $\conc_\eq$ to which the system relaxes depends on the initial conditions.
In ideal CRNs, there is a unique equilibrium state for each stoichiometric compatibility class, here labeled $\manifold{\big\{\consquantity^\unbrokenindex\big\}}$ (see Subs.~\ref{sub_equilibrium}).
In non-ideal CRNs, there might be multiple equilibria for every stoichiometric compatibility class.
Thus, we introduce $\manifold{\conc_\eq}\subseteq\manifold{\big\{\consquantity^\unbrokenindex\big\}}$ as the manifold in the concentration space such that any state in $\manifold{\conc_\eq}$ will relax towards the specific equilibrium $\conc_\eq$.
This implies that
\begin{equation}
\semigrand(\conc)\geq\semigrand(\conc_\eq)\,,\text{ }\text{ }\text{ }\forall \conc\in\manifold{\conc_\eq}\,.
\label{eq_g_manifold_bound}
\end{equation}

Let us stress two more points. 
First, the bound~\eqref{eq_g_manifold_bound} does not provide any constraint for the values of the semigrand free energy evaluated on different manifolds. 
Consequently, it is possible that $\exists \conc \notin\manifold{\conc_\eq}$ such that $\semigrand(\conc)<\semigrand(\conc_\eq)$.
Second, $\semigrand_\minl$ must be an equilibrium value of the semigrand free energy.
Indeed, if there was a nonequilibrium state $\conc$ such that $\semigrand(\conc)=\semigrand_\minl$, the relaxation of $\conc$ would necessarily correspond to a decrease of the semigrand free energy which is not possible because of Eq.~\eqref{eq_g_lowest_bound}.

We now turn to nonautonomous and/or nondetailed balanced CRNs.
The expression of the semigrand free energy~\eqref{eq_semi_grand_2} is exactly the same as for the corresponding autonomous detailed balanced CRNs. 
Thus, Eq.~\eqref{eq_g_manifold_bound} still holds with $\conc_\eq$ now the equilibrium state of the corresponding detailed balanced CRNs with $\setchemostatted = \setpotential$.
However, the evolution of $\conc(t)$ is constrained on a specific stoichiometric compatibility class $\manifold{\big\{\consquantity^\unbrokenindex\big\}}$, but not (necessarily) on a specific manifold $\manifold{\conc_\eq}$.
This means that different bounds~\eqref{eq_g_manifold_bound} can be identified during the same nondetailed balanced dynamics.

In conclusion, the semigrand free energy is always lower bounded by  $\semigrand(\conc_\eq)$ where $\conc_\eq$ is the equilibrium state to which the system would relax if it was detailed balanced. This, together with $\dt\semigrand(\conc(t)) = -T\epr(t)\leq0$ for autonomous detailed balanced CRNs, makes the semigrand free energy defined in Eq.~\eqref{eq_semigrand} the proper thermodynamic potential of open non-ideal CRNs. 
In other words, $\semigrand(\conc(t))-\semigrand(\conc_\eq)$ is a Lyapunov function.

\subsection{Relative Entropy\label{sub_relative_entropy}}
We show here that, because of the interactions, the difference $\semigrand(\conc(t))-\semigrand(\conc_\eq ) $ cannot be solely expressed as a relative entropy.
This is a significant difference with respect to previous studies on the thermodynamics of ideal CRNs~\cite{Rao2016, Rao2018b} and ideal reaction-diffusion systems~\cite{Falasco2018a, Avanzini2019a}.

To this aim, we notice that, because of the local detailed balance~\eqref{eq_local_detailed_balance_thermo}, the equilibrium chemical potentials satisfy
\begin{equation}
\boldsymbol\chempotential(\conc_\eq)\cdot\matS = 0\,.
\end{equation}
Thus, $\boldsymbol\chempotential(\conc_\eq)$ can be written as a linear combination of conservation laws~\eqref{eq_conservation_laws}, $\boldsymbol\chempotential(\conc_\eq)=\sum_\conslawindex\coeff_\conslawindex\boldsymbol\conslaw^\conslawindex$. 
Recalling that the conservation laws split into unbroken and broken conservation laws in open CRNs, we obtain 
\begin{equation}
\boldsymbol\chempotential(\conc_\eq)\cdot\conc_\eq  = \boldsymbol\chempotential(\conc_\eq)\cdot\conc(t) - \sum_\brokenindex\coeff_\brokenindex\boldsymbol\conslaw^\brokenindex \cdot\conc(t)+\sum_\brokenindex\coeff_\brokenindex\boldsymbol\conslaw^\brokenindex \cdot\conc_\eq\,.
\label{eq_thermo_pot_1}
\end{equation}
By recalling that the unbroken conservation laws are those with null entries for the chemostatted species, hence $\chempotential_\chemspecies(\conc_\eq)=\sum_\brokenindex\coeff_\brokenindex\conslaw^\brokenindex_\chemspecies$ for $\chemspecies \in \setchemostatted$, and the expressions for the concentrations of the moieties~\eqref{eq_moieties}, we verify that 
\begin{align}
&\boldsymbol\chempotential_\setpotential(t)\cdot\boldsymbol \moieties(\conc(t)) = \sum_\brokenindex\coeff_\brokenindex\boldsymbol\conslaw^\brokenindex \cdot\conc(t)\,,\label{eq_thermo_pot_2}\\
&\boldsymbol\chempotential_\setpotential(t)\cdot\boldsymbol \moieties(\conc_\eq) =\sum_\brokenindex\coeff_\brokenindex\boldsymbol\conslaw^\brokenindex \cdot\conc_\eq\,.\label{eq_thermo_pot_3}
\end{align}
By using Eq.~\eqref{eq_thermo_pot_1}, Eq.~\eqref{eq_thermo_pot_2}, and Eq.~\eqref{eq_thermo_pot_3}, we find that
\begin{equation}
\begin{split}
\semigrand(\conc(t))-\semigrand(\conc_\eq ) =& \left(\boldsymbol\chempotential(\conc(t))-\boldsymbol\chempotential(\conc_\eq)\right)\cdot\conc(t) +\\
&-RT (\norm{\conc(t)} - \norm{\conc_\eq})+\\
&+  \potenergy(\conc,\parameter)
-\boldsymbol\nabla_{\conc} \potenergy(\conc, \parameter)\cdot\conc+\\
&- \potenergy(\conc_\eq,\parameter)
+\boldsymbol\nabla_{\conc} \potenergy(\conc_\eq, \parameter)\cdot\conc_\eq\,.\\
\end{split}
\end{equation}
Finally, by using Eq.~\eqref{eq_elch_pot} to split the ideal and the non-ideal contribution of the chemical potential, we obtain 
\begin{equation}{
\begin{split}
\semigrand(\conc(t))-\semigrand(\conc_\eq )  =&RT\relentropy(\conc(t)\parallel\conc_\eq)+\\
&+\potenergy(\conc,\parameter) -  \potenergy(\conc_\eq,\parameter)+\\
&-\boldsymbol\nabla_{\conc} \potenergy(\conc_\eq, \parameter)\cdot(\conc-\conc_\eq)\,,
\end{split}}
\label{eq_thermo_pot_final}
\end{equation}
where we introduced the relative entropy for non-normalized concentration distributions
\begin{equation}
\relentropy(\boldsymbol a\parallel \boldsymbol b)= \sum_\muteindex a_\muteindex\ln\left(\frac{a_\muteindex}{b_\muteindex}\right)-(a_\muteindex-b_\muteindex)\geq0\,.
\end{equation}
Equation~\eqref{eq_thermo_pot_final} proves that $\semigrand(\conc(t))-\semigrand(\conc_\eq )$ cannot be written only as a relative entropy, but additional terms emerge from the interactions.
It is not granted that these additional terms are always positive unless $\potenergy(\conc, \parameter)$ is a convex function of $\conc$. 
Nevertheless, if $\conc(t)\in \manifold{\conc_\eq}$, $\semigrand(\conc(t))-\semigrand(\conc_\eq )\geq 0$ as proved in Subs.~\ref{sub_bound}.

\begin{example}
The difference $\semigrand(\conc)-\semigrand(\conc_\eq )$ in the CRN~\eqref{eq_crn_example}, where $\conc_\eq$ is the equilibrium to which the system would relax if only the $\setpotential$ species where chemostatted, can be written as 
\begin{equation}
\begin{split}
\semigrand(\conc)-\semigrand(\conc_\eq ) = &RT\relentropy(\conc(t)\parallel\conc_\eq)+\\
&+\frac{F\constant}{2}\big(\chargedensity_\compone(\conc,\parameter) - \chargedensity_\compone(\conc_\eq,\parameter)\big)^2+\\ 
&+\frac{F\constant}{2}\big(\chargedensity_\comptwo(\conc,\parameter) - \chargedensity_\comptwo(\conc_\eq,\parameter)\big)^2 \,.
\end{split}
\label{eq_crn_example_diff_g_geq}
\end{equation}
Since the interaction free energy~\eqref{eq_crn_example_int_pot_energ_tot} is a convex function, the terms in Eq.~\eqref{eq_crn_example_diff_g_geq} emerging from the electrostatic interactions are always positive, i.e., $\big(\chargedensity_{(\muteindex)}(\conc,\parameter) - \chargedensity_{(\muteindex)}(\conc_\eq,\parameter)\big)^2\geq 0$.
\qed
\end{example}
\begin{example}
We examine  the difference $\semigrand(\conc)-\semigrand(\conc_\eq )$ for the following closed CRN:
\begin{equation}
\ch{B <=>[ ${+1}$ ][ ${-1}$ ] A^{-} +  C^{+}}\,,
\label{eq_crn_example3}
\end{equation}
when the electrostatic interactions are described within Debye-H\"uckel theory~\cite{Mcquarrie2000}:
 \begin{equation}
\potenergy(\conc) = -\frac{4F\costone}{\costtwo^3}\bigg\{\frac{\costtwo^2}{2}\mathcal I(\conc)  - \costtwo\sqrt{\mathcal I(\conc) } + \ln\Big[\costtwo\sqrt{\mathcal I(\conc)} + 1\Big]\bigg\}\,,
\end{equation}
where $\mathcal I(\conc) $ is the ionic strength 
\begin{equation}
\mathcal I(\conc) = \frac{1}{2}([\ch{A^{-}}]+[\ch{C^{+}}])\geq0\,,
\end{equation}
and $\costone$ and $\costtwo$ are positive constants depending on the specific system (see Ref.~\onlinecite{Mcquarrie2000} for their explicit expressions).
In this case  $\semigrand(\conc)-\semigrand(\conc_\eq )$  becomes
\begin{equation}
\begin{split}
\semigrand(\conc)-\semigrand(\conc_\eq ) = &RT\relentropy(\conc(t)\parallel\conc_\eq)+\\
&+ \big(\potenergy(\conc) - \potenergy(\conc_\eq)\big) +\\
& + 2 F \costone \big[\mathcal I(\conc) - \mathcal I(\conc_\eq)\big]\frac{\sqrt{\mathcal I(\conc_\eq)}}{1+\costtwo\sqrt{\mathcal I(\conc_\eq)}}\,.
\end{split}
\label{eq_diff_DH}
\end{equation}
Here the terms emerging from the electrostatic interactions may be negative because the interaction free energy provided by the Debye-H\"uckel theory is not a convex function.
\qed
\end{example}


\section{Conclusions\label{sec_conclusion}}
In this work, we developed a thermodynamic theory of non-ideal CRNs.
We used activity coefficients to account for interactions within a mean-field approach.
Our framework is general and holds for any (thermodynamic consistent) expressions of the interactions.
Exploiting the local detailed balance assumption, we showed that the dynamics of non-ideal CRNs does not follow mass-action kinetics unlike for ideal-CRNs.
Generalizing some results developed for ideal-CRNs, we used conservation laws and cycles to derive physically meaningful decompositions of the entropy production rate.
The first~\eqref{eq_epr_decomposition} is derived exploiting the conservation laws and allowed us to determine the proper thermodynamic potential and the forces driving non-ideal CRNs out of equilibrium.
The second~\eqref{eq_epr_decomposition2} is derived exploiting the cycles and allowed us to express the steady-state entropy production using only the chemical potential of the chemostatted species and the currents along the emergent cycles.
We proved that the thermodynamic potential acts as a Lyapunov function in detailed balanced CRNs.

Crucially, we showed  that the thermodynamic structure of non-ideal CRNs is the same as in ideal CRNs.
In particular, the conservation laws and cycles determine whether systems are detailed balance or not
and define the forces~\eqref{eq_nc_force} maintaining nonequilibrium regimes in the same way for ideal and non-ideal CRNs.

These results are fundamental to characterize processes of energy transduction in biosystems and electrochemical systems where interacting ions are ubiquitous.
As a perspective, one could exploit the same strategy we followed in this work to specialize the stochastic thermodynamics of CRNs to interacting systems.
We leave these points to future investigations.

\section{Acknowledgments}
This research was funded by the European Research Council project NanoThermo (ERC-2015-CoG Agreement No.~681456).



\bibliography{biblio}
\end{document}